\documentclass{article}
\pdfoutput=1
\usepackage{algorithmic}
\usepackage{algorithm}
\usepackage{amssymb,amsmath,amsfonts}
\usepackage{graphicx}
\usepackage{verbatim}
\usepackage{subfigure}
\usepackage{url}
\usepackage{empheq}
\usepackage[final]{changes}
\usepackage{changes}
\usepackage{tikz}
\usetikzlibrary{positioning,chains,fit,shapes,calc,arrows,automata}

\newtheorem{defi}{Definition}

\newcommand{\U}{\mathcal{U}}
\newcommand{\A}{\mathcal{A}}
\newcommand{\D}{\mathcal{D}}
\newcommand{\N}{\mathcal{N}}
\newcommand{\T}{\mathcal{T}}
\newcommand{\Ao}{\A^o}
\newcommand{\At}{\A^{24}}
\newcommand{\Uo}{\U^o}
\newcommand{\Ut}{\U^{24}}
\newcommand{\Utb}{\mathcal{U}_b^{24}}
\newcommand{\Uob}{\mathcal{U}_b^{o}}
\newcommand{\Atb}{\mathcal{A}_b^{24}}
\newcommand{\Aob}{\mathcal{A}_b^{o}}

\begin{document}

\title{Designing Low Cost and Energy Efficient Access Network for the Developing World}

\author{
  Kshitiz Verma\\ IIT Kanpur,India \\ UC3M, Spain\\
  \texttt{vermasharp@gmail.com}
  \and
  Shmuel Zaks \\ Technion, Haifa, Israel\\
  \texttt{zaks@cs.technion.ac.il}
  \and
  Alberto Garcia-Martinez \\ UC3M, Spain\\
  \texttt{alberto@it.uc3m.es}
}


\maketitle

\begin{abstract}
Internet is growing rapidly in the developing world now. Our survey of four networks in India, all having at least one thousand users, suggest that both installation cost and recurring cost due to power consumption pose a challenge in its deployment in developing countries. In this paper, we first model the access design problem by dividing the users in two types 1) those that may access the network anytime and 2) those who need it only during office hours on working days. The problem is formulated as a binary integer linear program which turns out to be NP-hard. We then give a distributed heuristic for network design. We evaluate our model and heuristic using real data collected from IIT Kanpur LAN for more than 50 days. Results show that even in a tree topology -- which is a common characteristic of all networks who participated in our study, our design can reduce the energy consumption of the network by up to 11\% in residential-cum-office environments and up to 22\% in office-only environments in comparison with current methods without giving up on the performance. The extra cost incurred due to our design can be compensated in less than an year by saving in electricity bill of the network.
\end{abstract}

\section{Introduction} 
\label{sec:intro}

\added{\subsection{Networks in developing countries} }
The Government of India recently announced its \$1 billion project to build an internal National Knowledge Network (NKN) to connect 1500 institutions within India \cite{nkn}. Letters were sent to colleges to join NKN and setting up a LAN with government funding \cite{sakshat}. While some of the colleges receive funding from the government, not all the institutions and organizations have such funding available. So installing a network remains challenging for most of the universities as well as other organizations because networking equipments are expensive.

Apart from the installation cost, all networks have a periodic operational cost and most of which is due to the 24 hour electricity consumption of networking infrastructure. Hence, besides the need for design to be cost efficient to install, it should also support switching off networking devices whenever electricity can be saved. However, it is extremely important to understand that the networks deployed depend on the requirement and expectations of the users subject to the availability of necessary resources. It is important the implementations in developing countries are low cost \cite{james2002low}. One cannot simply copy what was successfully implemented in developed countries to the developing countries \cite{osin1998computers}\cite{brewer2005case}. For example, the network upgrade of the University of Texas at Austin during 2011-2012 had an estimated replacement cost of \$60-\$65 million \cite{austinreport11}, which is equivalent to three times the yearly public funding to even the best universities in India \cite{iitkfunding}. Thus, shortness of budget has fundamental effect on design of networks and prevents keeping redundant nodes and links, resulting in networks that are very sensitive to failures. Undoubtedly, cost is the biggest retarding factor in deployment of networks in India (particularly universities), and power consumption plays the next important role.


\begin{figure*}[t]
\begin{minipage}[b]{0.48\linewidth}\centering
\includegraphics[width=\textwidth]{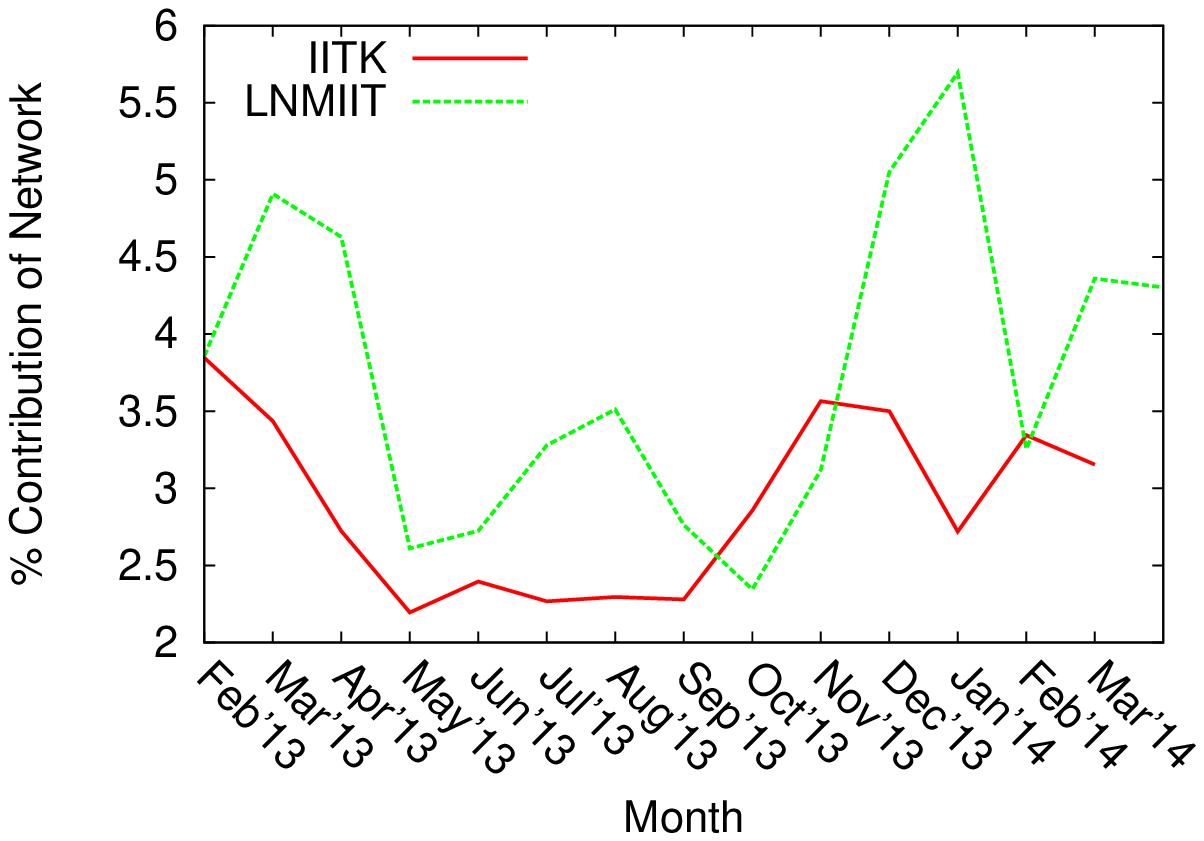}
\caption{Network power consumption as a percentage of total consumption as per the data obtained from competent authorities. Note that IITK Network is ten times greater than LNMIIT Network, both in terms of users and diameter.}\label{fig:powerconsumption}
\end{minipage}
\hspace{3mm}
\begin{minipage}[b]{0.48\linewidth}\centering
\includegraphics[width=\textwidth]{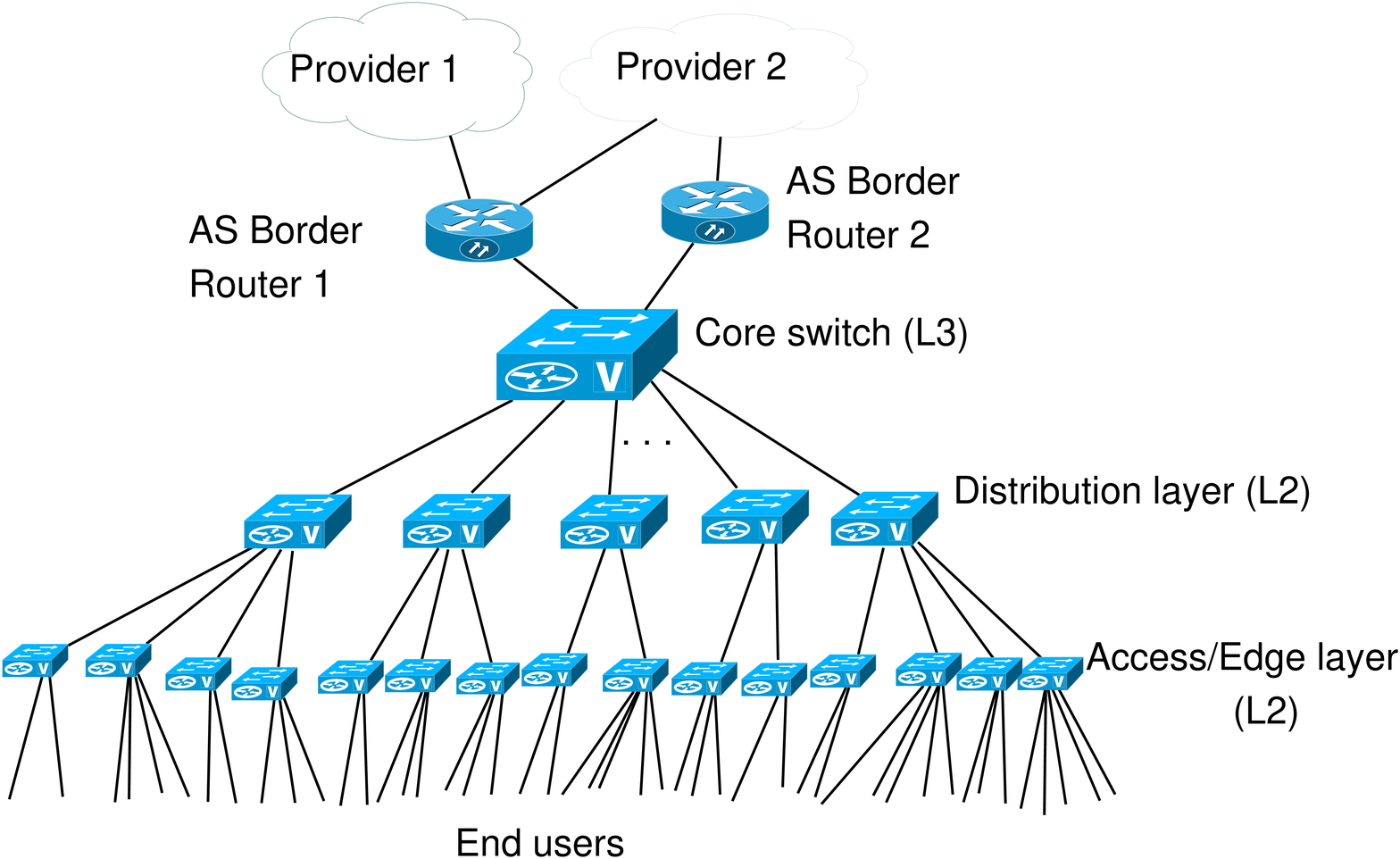}
\caption{Typical hierarchical tree topology of networks in India. The routers are kept only at the edge towards ISP and topology below the core switch is a common characteristic of four networks that shared this information.}\label{fig:tree}
\end{minipage}
\hspace{3mm}
\end{figure*}

Fig. \ref{fig:powerconsumption} shows how significant the power consumed by the network is. It plots the sum of power consumption of all the networking devices (e.g., switches, routers, UPS backup for switches, firewalls, etc.) in IIT Kanpur(IITK) and the LNMIIT Jaipur Network, as a percentage of the total power consumption of the corresponding university. However, it does not include the power consumption of the end terminals. \deleted{On similar tracks, power consumption of Network-2, which is deployed in another university, is also shown as a percentage of the total power consumption of the university.} The maximum percentage reaches as high as 5.5! The two curves exhibit similar properties, network power consumption percentage is low during summer. It is due to heavy use of air conditioning as the temperature in both the cities rise above $45\,^{\circ}\mathrm{C}$, whereas the network power consumption stays constant. During winters, heating is not used and ceiling fans in the buildings are also kept switched off. Hence, the power consumption percentage of the network rises. 

Interestingly, as we will see later, even the temperature plays a big role in the design of networks. Choosing correct vendors and switches may help in designing more robust networks.

As a result of our survey of four different networks, in different regions of India, all serving at least one thousand users, it turns out that there are no redundant links or nodes in Indian networks. The topology is hierarchical with core switch at the root of the intranet switches which connects the network to the Internet by connecting to routers via a firewall (Fig. \ref{fig:tree}). Any two nodes in the intra-network have exactly one path between them. If there is any failure in the network, the Computer Center staff fixes the problem by visiting the site of failure. Until the problem is fixed, all the users served by the failed node/link remain disconnected from the network. Note that the absence of routers from the topology. It is because they are expensive and implementing static routing is far cheaper than using dynamic protocols like OSPF or RIP. In IITK-Network, all the distribution and edge layer switches are managed L2 switches with the core switch as the only L3 switch. 

Our work provides a first view on the networks specific to the developing world. Lack of a clear methodology to design network have led to inefficient networks like the one at the LNMIIT Jaipur. The university is now considering redesign of the whole network. It was expanded in an adhoc manner and the current situation is such that the Computer Center staff cannot configure all the switches centrally. They have to visit the switches themselves. If there is any failure, they must try to figure out the malfunctioning switch manually by visiting a set of possible switches using hit and trial.

\added{\subsection{Energy efficient design}}
Energy requirements and availability are different as well. The cost of electricity in India is no lesser than four times the cost of electricity in any of the developed countries whereas funding available to the institutions can be up to 4 orders of magnitude smaller. 

In this paper, we focus our attention on design of access networks in developing economies which have different requirements and expectations from their counterparts in the developed world. In particular, we studied one of the biggest LANs (IITK-Network) in India. Apart from detailed switch level topology, we collected \deleted{various networking parameters like} interface summary every five minutes for each core link and every 20 minutes for user links for more than 50 days. For the same time period, we also collect the data as a result of pinging all the switches in the network from one machine. The goal is to study the properties and utilization of the network and to see if even in a tree topology it is possible to switch off nodes during night.


\begin{table}[t]
\caption{Cable types used in IITK-Network.}
\label{tab:n1}
\begin{center}
\begin{tabular*}{98.3mm}{|c|c|p{27mm}|}
 \hline
 \textbf{Link type} & \textbf{Fiber/Cable used} & \textbf{Link Capacity} \\
 \hline\hline
 Core-Distribution & Optical Fiber & 1 Gbps\\ \hline
 Distribution-Edge & Copper & 1 Gbps\\ \hline
 Edge-User &  Copper & 100 Mbps\\ \hline
\end{tabular*}
\end{center}
\normalsize
\end{table}

\begin{figure}[h!]
\centering
\includegraphics[width=8cm]{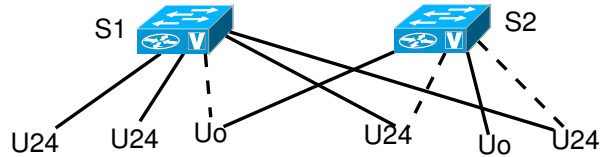}
\caption{The length of the lines represent the actual distance between the users and switches. A user has two lines emerging if it is connected to a switch other than the closest one. Dashed lines represent the connection to the closest switch, the current practice. Solid lines represent actual connection in our design. All U24 are connected to S1 and all Uo to S2. }
\label{fig:idea}
\end{figure}

To achieve this, we divide users of the network in two groups that, 1) need access to the network round the clock (type U24) and 2) access the network only during office hours (type Uo). In the current design, there is no distinction between the users and each user is connected to the closest available switch in its vicinity. We propose that all users of type Uo connect to a switch that does not have any user of type U24 and vice versa. This may result in a connection to a switch that may not be the closest available switch, potentially increasing the length of the cables in the network (Fig. \ref{fig:idea}).

\subsection{Our Contribution}
\label{subsec:ourcontri}
The question that we answer in this paper is: Can we connect end users to access switches in a manner so that the resulting network supports switching off some of the access switches when not in use? We answer this question positively. Our major contributions are as follows:

\begin{enumerate}
\item We provide a systematic method for designing access networks for developing countries. We model the problem as a binary linear integer program and show that designing network with optimal cost and power consumption is NP-hard. Hence, we provide a tractable heuristic. 
\item We observe through real data that some users indeed leave the network during night hours and never use it until morning. 
\item We evaluate our model using real data from an Indian network having more than ten thousand users to conclude that up to 11\% reduction in power consumption is possible even when the network has tree topology. 
\item We identify ambient temperature as one of the major design aspects for networks in developing countries. We provide a comparison of switches which can withstand high temperature.

\end{enumerate}

The rest of the paper is organized as follows: Section \ref{sec:model} gives a glimpse of the detailed model and problem formulation. In Section \ref{sec:approach}, we describe our approach to energy saving. Section \ref{sec:evaluation} provides a comprehensive evaluation of our approach using the real data collected from IITK-Network. Section \ref{sec:pinging} describes the results of pinging the switches in the network. Section \ref{sec:relatedwork} discusses the related work in the area and finally we conclude in Section \ref{sec:conclusion}.

\section{Model}
\label{sec:model}

\deleted{The model presented here is specific to cheap networks that are typically suited to the developing world.} The network is modeled as switches with different levels with a cost associated to the nodes and links. Power consumption of the network is modeled as the total power consumed by the network devices. Finally, users are modeled in two categories to implement energy savings.

\subsection{Network Model}
\label{subsec:netmodel}
We consider a network $\N$ that contains users $\U =\{u_1,u_2, \cdots, u_U\}$, access switches $\A=\{a_1,a_2, \cdots, a_A\}$, distribution switches $\D = \{d_1,d_2, \cdots, d_D\}$, and one core switch $s$.
Every user $u \in\U$ is connected to every access switch $a \in\A$ by a link of capacity $C_{ua}$ and length $l_{ua}$. Every access switch $a \in\A$ is connected to every distribution switch $d \in\D$ by a link of capacity $C_{ad}$ and length $l_{ad}$.  Every distribution switch $d \in\D$ is  connected to the core switch $s$ by a link of capacity $C_{ds}$ and length $l_{ds}$. \deleted{(Actually, the links between users and access switches and between access switches and distribution switches are of the same type $\tau_{copper}$, while those between switches and the core are of another type $\tau_{fiber}$, but these are reflected in the corresponding costs.)} 

\deleted {Within this network we are looking for a sub-network $\T$ of a tree structure, that will connect all users in \U to the core switch $s$. That is, the tree nodes are composed of all users and the core switch $s$, and of a subset of the access switches $\A$ and distribution switches $\D$. The access switch to which user $u$ is connected will be denoted by $a_u$. In addition, it is required that in $\T$ every access switch $a$ is connected to at most $\delta_a$ users, and every distribution switch $d$ is connected to at most $\delta_d$ access switches. (We assume that the core switch can accommodate large number of distribution switches and its capacity can be increased whenever required.)
Thus, in the required sub-network $\T$ there is no link redundancy, i.e., any two nodes have exactly one path between them; in other words, if any of the nodes or links fails, the tree $\T$ is partitioned into two or more disconnected fragments.
The sub-network $\T$  is chosen according to the cost model described below.}

\subsection{Cost Model}
\label{subsec:cm}
In our settings, the routers are present only at the edge towards ISP and are necessarily on. Hence, we consider only switches in our model and ignore routers. 
\subsubsection{Establishment cost}
\label{subsubsec:est}

With each node or switch $v$ and each link $(v,w)$ of the network we associate costs $c_v$ and $c_{vw}$, correspondingly.  Specifically, for every user $u$, access switch $a$, distribution switch $d$, the costs of $u$, $a$, $d$, and the core switch $s$ are $c_u$, $c_a$, $c_d$ and $c_s$, and the costs of links $(u,a)$, $(a,d)$, and $(d,s)$ are $c_{ua}$,  $c_{ad}$, and $c_{ds}$, correspondingly. Since networks are being planned for users, we assume that there is no cost associated to any user
(i.e.,  $c_u=0$ for every $u \in\U$).
\deleted{Note that this cost depends upon the length and the type of the cable.} This cost is paid once when the network is established. So we call it the \emph{establishment cost}.

\subsubsection{Operational cost}
\label{subsubsec:op}

\deleted{We consider energy consumed by the networking devices only, i.e., all switches. As mentioned before, intranet consists of only switches, so we consider a network of only switches and ignore routers in our model.} The power consumption of switch $v$ is denoted by $P_v$; specifically, the power consumptions of the core switch $s$, of every distribution switch
$d$, and of every access switch $a$, is denoted by $P_s$,  $P_d$,  and $P_{a}$, correspondingly.
We assume that the power consumed by
the network cables is negligible compared to the power consumed by the devices. We also assume that the devices are not energy proportional, i.e., they consume the same power when idle or when in working state. This is true for networks in developing countries because the switches used are still legacy as either the new energy efficient hardware has not been deployed or is too costly to deploy.

Thus, the power consumption $P$ of the network built above is given by,
\begin{equation}
P = P_s + \sum_{d=1}^{D} P_d + \sum_{a=1}^{A} P_{a}
\label{eq:power}
\end{equation}
The power consumed by the network has to be paid monthly, and is thus termed as the \emph{operational cost}.

\begin{figure}[t]
\centering
\definecolor{myblue}{RGB}{80,80,160}
\definecolor{mygreen}{RGB}{80,160,80}

\begin{tikzpicture}[thick,
  fsnode/.style={draw,circle,fill=myblue},
  ssnode/.style={draw,circle,fill=mygreen},
  every fit/.style={inner sep=-2pt,text width=2cm}
]

\begin{scope}[start chain=going below,node distance=4mm]
\foreach \i in {1,2,3,4,5,6}
  \node[fsnode,on chain] (f\i) [label=left: \i] {};
\end{scope}

\begin{scope}[xshift=3cm,yshift=-0.8cm,start chain=going below,node
distance=7mm]
\foreach \i in {1,2,3}
  \node[ssnode,on chain] (s\i) [label=right: \i] {$c_\i$};
\end{scope}

\node [myblue,fit={(f1) (f6)},label=left:Users] {};
\node [mygreen,fit=(s1) (s3),label=right:Access Switches] {};

\draw (f1) -- (s1) node[near start, above] {$c_{11},l_{11}$};
\draw (f2) -- (s1) node[near start, above] {$c_{21},l_{21}$};
\draw (f3) -- (s1) node[near start, above] {$c_{31},l_{31}$};
\draw (f4) -- (s2) node[near start, above] {$c_{42},l_{42}$};
\draw (f5) -- (s2) node[near start, above] {$c_{52},l_{52}$};
\draw (f6) -- (s3) node[near start, above] {$c_{63},l_{63}$};

\end{tikzpicture}
\caption{A graph to represent the edge of the network. Each user $u$ is allocated to exactly one access switch $a$. Whereas a switch $a$ can be connected to at most $\delta_a$ users. The cost of the switches is written in the nodes and the cost and the length of the links is written above them. Cost of all users is 0.}
\label{fig:bm}
\end{figure}
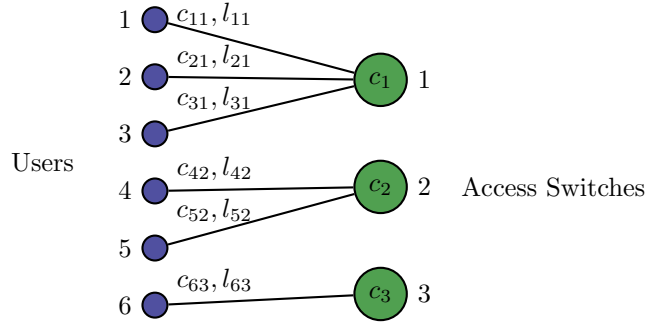

\begin{table}[t]
\caption{Notation used in this work.}\label{tab:notation}
\begin{center}
\begin{tabular*}{80mm}{|c|p{58.3mm}|}
 \hline
 \textbf{Symbol} & \textbf{Definition} \\
 \hline\hline
 $\N$ & the given network \\ \hline
 $\T$ & the sub-network of a tree topology\\ \hline
 $\U$ & the set of users in the network \\ \hline
 $s$ & the core switch \\ \hline
 $u$ & a user  \\ \hline
 $a$ & an access switch \\ \hline
 $d$ & a distribution switch \\ \hline
 $(i,j)$ & Link between nodes $i$ and $j$ \\ \hline
 $C_{ij}$ & Bandwidth of link $(i,j)$\\ \hline
 $l_{ij}$ & Length of link $(i,j)$  \\ \hline
 $c_{ij}$ & Cost of link $(i,j)$ \\ \hline
 $c_i$ & Cost of node $i$ \\ \hline
 $P_i$ & Power consumed by node $i$ \\ \hline
 $\delta_i$ & Maximum degree of node $i$ \\ \hline
\end{tabular*}
\end{center}
\normalsize
\end{table}

\subsection{User and Access Switch Profiles}
\label{subsec:userp}
Energy savings depend upon user profiles. Hence, we need to study the user profiles of the network to switch off the devices in the network. \deleted{Let the set of all end users be $\U$.} To achieve this, we further divide the end user set $\U$ into two categories:
\begin{enumerate}
\item $\Uo$: The set of users that are going to access the network only during office hours. Office hours can be set as desired.
\item $\Ut$: The set of users who may access the network anytime. Hence, it is not desirable to turn off the switches that are attached to the
users of this type.
\end{enumerate}

Specifically, we impose the following restriction on all end users.
\begin{itemize}
\item $\Uo \cup \Ut = \U$
\item $\Uo \cap \Ut = \emptyset$
\end{itemize}

These two conditions together imply that a user belongs to exactly one of the two categories. Note that in the definition of $\Uo$ users, it is implicitly assumed that they never use their PCs once they leave the workplace. We observe this for the users in real data that is discussed in Section \ref{sec:evaluation}.

As with the users, we divide the access switches in similar two categories, $\Ao$ and $\At$. Again for them,
\begin{itemize}
\item $\Ao \cup \At = \A$
\item $\Ao \cap \At = \emptyset$
\end{itemize}

\begin{figure}[t]
\centering
\subfigure[During day all the switches are ON]
{
\includegraphics[width=7cm]{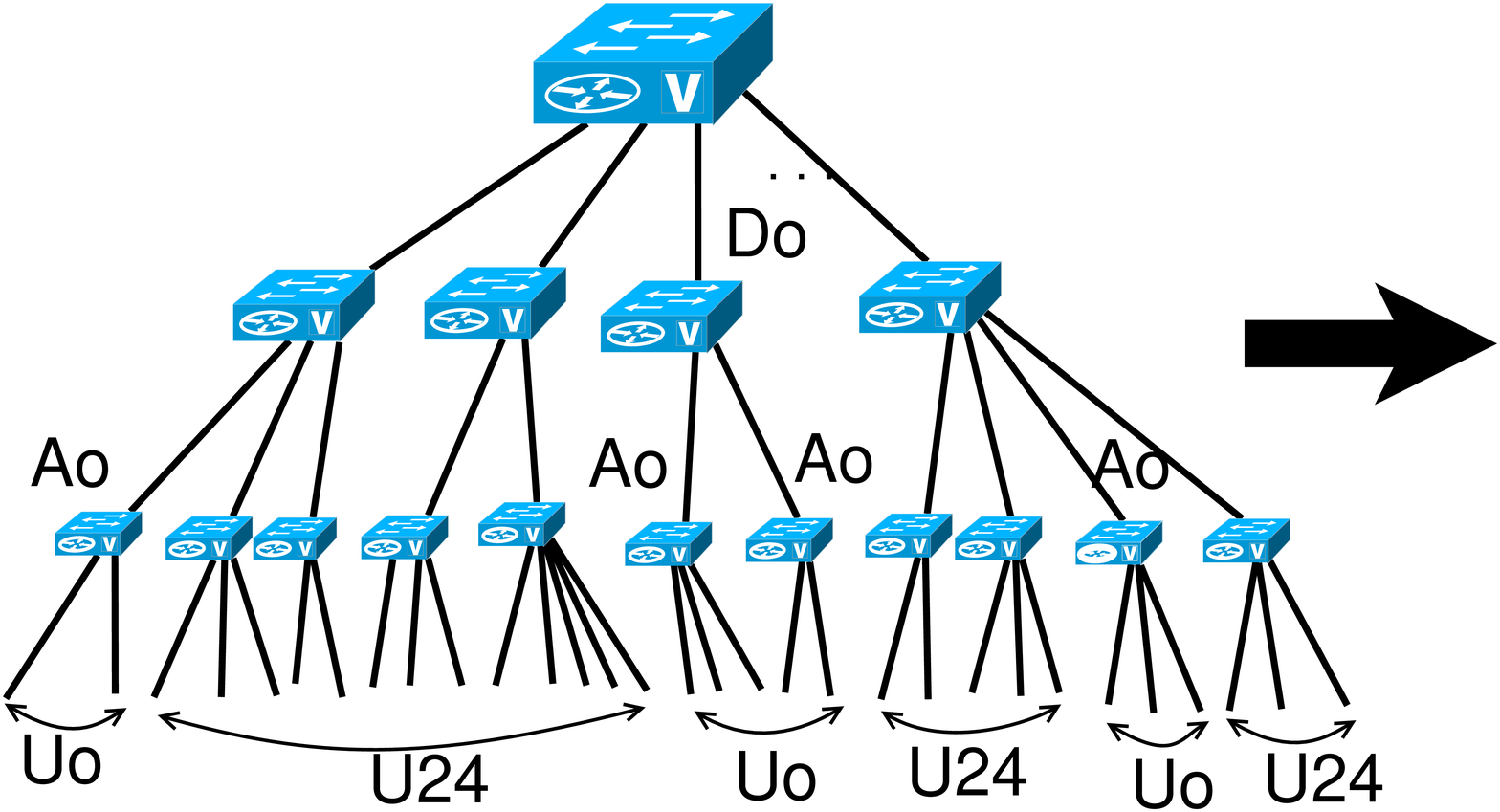}
\label{fig:during_day}
}
\subfigure[During night $A_o\in\Ao$ and $D_o$ can be turned off]
{
\includegraphics[width=7cm]{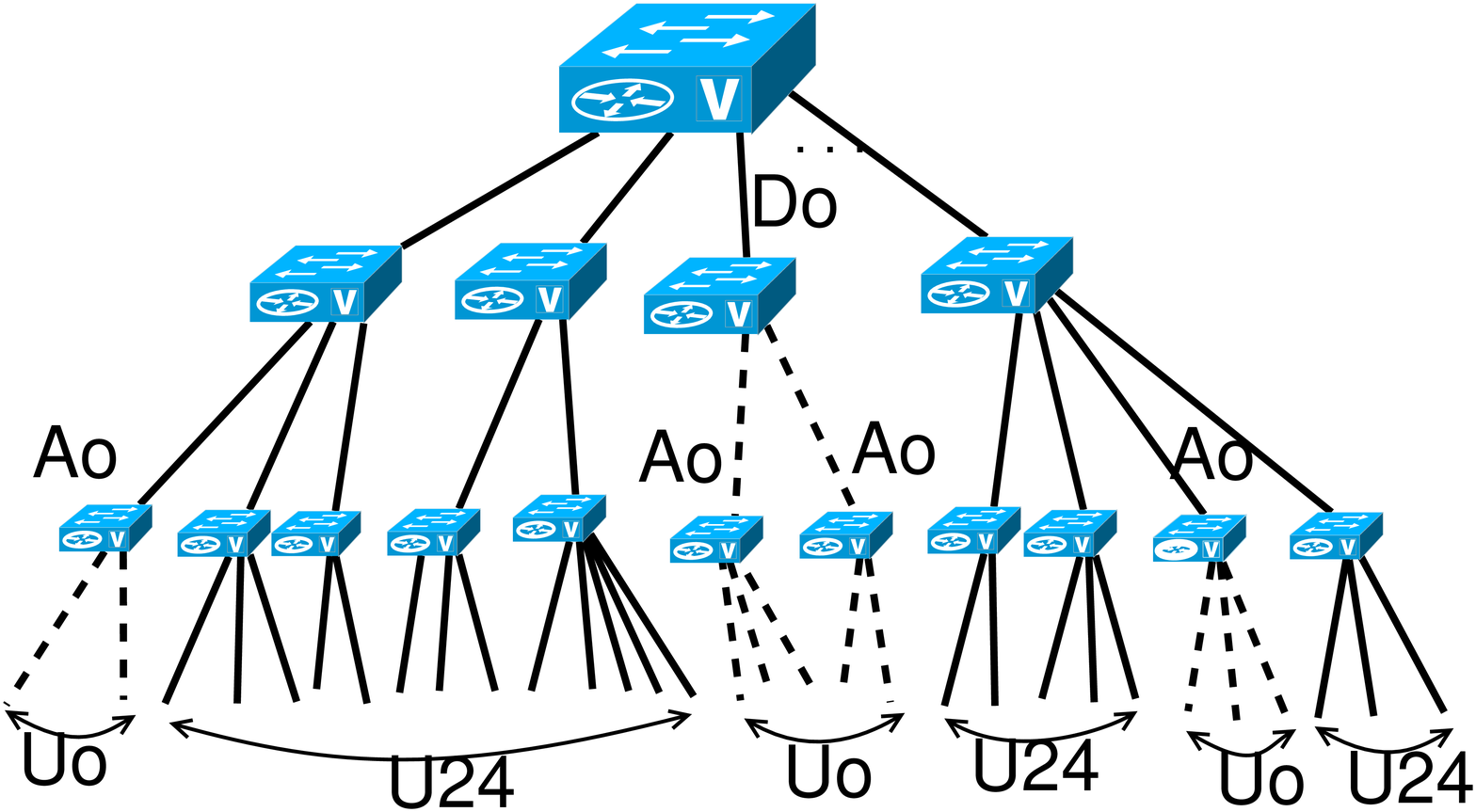}
\label{fig:during_night}
}
\caption{Basic idea behind the energy saving network design. $A_o\in \Ao$ switches can be turned off because all the users connected to them are not present in the night. $D_o$ can be turned off because all switches connected to it are of type $\Ao$.}
\label{fig:approach}
\end{figure}

We impose conditions that $\Uo$ can be \replaced{connected }{associated} only to $\Ao$ and $\Ut$ only to $\At$. \deleted{ as long as 
the distance criteria are satisfied. To be specific, we connect all the users in a building to only the 
access switches in the same building. We keep the different profiles of switches only until access 
switches. Hence, } Distribution switches are only of one kind and both the kinds of access switch may connect 
to it. However, a distribution switch can be switched off if all the switches connected to it are of 
type $\Ao$ \added{and are all switched off.} Fig. \ref{fig:approach} describes the underlying idea.

\subsection{Problem and its complexity}
\label{subsec:problemformulation}

\noindent {\bf Problem CHEAP-NETWORK:}

\noindent {\it Input:} A network $\mathcal{N}$, with users $\U= \Uo \cup \Ut = \{1,2, \cdots, \mathcal{U}\}$, $\Uo \cap \Ut = \emptyset$,  access switches $\A= \Ao \cup \At = \{1,2, \cdots, \mathcal{A}\}$, $\Ao \cap \At = \emptyset$, distribution switches $\D=\{1,2, \cdots, \mathcal{D}\}$, and a core switch $s$. Every user $u \in\U$ is connected to every access switch $a \in\A$ by a link of capacity $C_{ua}$ \deleted{type $\tau_{copper}$} and length $l_{ua}$. Every access switch $a \in\A$ is connected to every distribution switch $d \in\D$ by a link of capacity $C_{ad}$ \deleted{type $\tau_{copper}$} and length $l_{ad}$.  Every distribution switch $d \in\D$ is  connected to the core switch $s$ by a link of capacity $C_{ds}$ \deleted{ type $\tau_{fiber}$} and length $l_{ds}$. 
A bound $\delta_a$ for each access switch $a \in\A$, and a bound $\delta_d$  for each distribution switch $d \in\D$. \deleted{ , and $L,$ $\alpha\geq 1,\beta\geq 0$. All parameters are non-negative real numbers.}

\noindent {\it Output:} A sub-network $\mathcal{T}$ of $\mathcal{N}$, \deleted{ of a tree topology,} that includes all users in $\U$ and the core switch $s$, in which a user in $\Uo$
 ($\Ut$) is connected to an access switch in $\Ao$ ($\At$).

\noindent {\it Objective:} Minimize the total \added{establishment} cost. \deleted{ given by the sum of the establishment cost and the operational cost, and such that the total wire length used is bounded by $L$.}

Problem CHEAP-NETWORK can be formalized as the following binary integer linear program.\\

Minimize,
\begin{eqnarray}
& & \sum_{a=1}^{A}\sum_{u=1}^{\mathcal{U}}c_{ua}\cdot w_{ua} + \sum_{a=1}^{A}c_a\cdot x_a + \sum_{d=1}^{D}\sum_{a=1}^{A}c_{ad}\cdot y_{ad} \nonumber\\
 & & +\sum_{d=1}^{D}(c_{d}+c_{ds})\cdot z_d +c_s
\label{eq:cost}
\end{eqnarray}

subject to,

\begin{empheq}[left=\text{Edge layer}\empheqlbrace]{align}
\sum_{a \in \Ao} w_{ua} & = 1, ~~~~\forall u\in \Uo  \label{eq:c1a} \\
\sum_{a \in \At} w_{ua} & = 1, ~~~~\forall u\in \Ut   \label{eq:c1b} \\
\sum_{u=1}^\mathcal{U} w_{ua} & \leq \delta_a, ~~~~\forall a\in \A \label{eq:c2} \\
w_{ua} & \leq x_a, ~~~~\forall u\in \U , a\in \A \label{eq:c3}
\end{empheq}
\begin{empheq}[left=\text{Distribution layer}\empheqlbrace]{align}
\sum_{d=1}^D y_{ad} & = x_a, ~~~~\forall a\in \A  \label{eq:c4} \\
\sum_{a=1}^A y_{ad} & \leq  \delta_d, ~~~~\forall d\in \D \label{eq:c5}\\
y_{ad}  & \leq z_d, ~~~~\forall a\in \A , d\in \D \label{eq:c6}
\end{empheq}

where,

\begin{equation}
\left.
\begin{aligned}
w_{ua} \in \{ 0,1 \} ~~~~ \forall u \in \U, \forall a \in \A \\ \nonumber
y_{ad} \in \{ 0,1 \} ~~~~ \forall a \in \A, \forall d \in \D \\ \nonumber
x_a \in \{ 0,1 \} ~~~~ \forall a \in \A \\ \nonumber
z_d \in \{ 0,1 \} ~~~~ \forall d \in \D \nonumber       
\end{aligned}
~~\right\}
\text{Binary variables}
\end{equation}

The variable $x_a$ ($z_d$) is $1$ or $0$ depending on whether access switch $a$ (distribution switch $d$) is chosen or not.
The variable $w_{ua}$ ($y_{ad}$) is set to $1$ or $0$ depending on whether user $u$ (access switch $a$)
is connected in the solution $\mathcal{T}$ to access switch $a$ (distribution switch $d$). Eqs. \ref{eq:c1a} and \ref{eq:c1b} state that each user must be connected to exactly one access switch of the corresponding type. Eq. \ref{eq:c2} implies that at most $\delta_a$ users may connect to access switch $a$. Eq. \ref{eq:c3} ensures that an edge between user $u$ and access switch $a$ is possible only if $a$ is chosen to be installed.
Eq. \ref{eq:c4} states that each chosen access switch must be connected to at most one distribution switch.
\deleted{(and otherwise to no distribution switch)} Eq. \ref{eq:c5} puts a constraint on the degree of a
distribution switch. Eq. \ref{eq:c6} states that an edge between access switch $a$ and distribution switch
$d$ is possible only if $d$ is chosen to be installed.
\deleted{Eq. \ref{eq:c7} states the the total wire length connecting user $u$ to an access switch is bounded by a function of the minimum possible length connecting $u$ to any access switch in $\A$ (recall that $a_u$, used in Eq. \ref{eq:c7}, is a unique access switch in $\T$ connected to user $u$).}

The above formulation of cost minimization problem belongs to the class of Facility Location Problem, which is known to be NP-hard \cite{andrews1998access}. Fig. \ref{fig:bm} shows an instance of the problem at the access layer. It can be easily seen that the solution $\T$ provided by the above formulation is a tree with core switch being at the root.

\section{Heuristic}
\label{sec:approach}

Since the problem is NP-hard and the number of users in a network can be in tens of thousands, it is infeasible to solve it optimally. Moreover, the design should allow expansion of the network in future. Hence, it is better that the network should be built in a distributed manner. For example, it can be built for different buildings and then each network of the building should be connected to the core switch, which is maintained centrally. Each building can be kept as a different virtual LAN (VLAN), separating the link layer broadcast domains. In each building, dedicated locations for keeping switches should be made. In general, buildings have smaller area, and keeping switches together at one place is far more desirable than keeping each switch in a different location. \deleted{Hence, $\alpha$ and $\beta$ in Eq. \ref{eq:c7} are chosen to suit this requirement.} The sites for keeping switches should be inside the building. As we will see later, ambient temperature can influence the network if switches are kept outside the buildings.

\begin{algorithm}[h!]
\caption{Heuristic for network design}
\label{algo1}
\begin{algorithmic}[1]
\STATE Let the set of all buildings be $\mathcal{B}$
\STATE For every building $b\in \mathcal{B}$, the set of users are given by ${\U}_b$, and set of access switches by $\A_b$, satisfying
\begin{equation}
\U = \bigcup_{b\in \mathcal{B}} {\U}_b
\end{equation}
\begin{equation}
\forall~ i,j\in \mathcal{B}, ~~~{\U}_i \cap {\U}_j = \emptyset, ~~ i\neq j
\end{equation}
\STATE Identify potential sites in a building where access switches can be installed. The site for switches should be well protected from outside temperature.
\STATE Divide users in building $b$ in $\Uob$ and $\Utb$, such that $\Uob \cup \Utb = \U_b$ and $\Uob \cap \Utb = \emptyset$  
\STATE Divide access switches in building $b$ in $\Aob$ and $\Atb$, such that $\Aob \cup \Atb = \A_b$ and $\Aob \cap \Atb = \emptyset$ 
\STATE All users in $\Uob$ are connected to the closest $\Aob$, and all the users in $\Utb$ are connected to the closest $\Atb$
\STATE Every building $b$ should have one distribution switch which is connected to the core switch. All $\Aob$ and $\Atb$ are connected to the distribution switch. 
\end{algorithmic}
\end{algorithm}

The ILP formation mentioned before minimizes the cost of the whole network and hence we call it a centralized design. On the other hand, the heuristic presented in Algorithm \ref{algo1} aims to minimize the number of switches in the building but this may result in increase in the length of total wire used for a building. We do a cost analysis of this as well.

\section{Evaluation}
\label{sec:evaluation}

We evaluate our framework using real data collected for more than 50 days from India's one of the largest access network having close to ten thousand end user terminals. Our major finding is the existence of users that do not access the network during night, \deleted {a fixed defined time, during night, to be specific,} allowing turning off the access switches even in a tree topology. This results in reducing the power consumption of the network by up to 11\% at IIT Kanpur. 

\subsection{Data Collection}
\label{subsec:dc}
\begin{figure*}[t]
\begin{minipage}[b]{0.48\linewidth}\centering
\includegraphics[width=\textwidth]{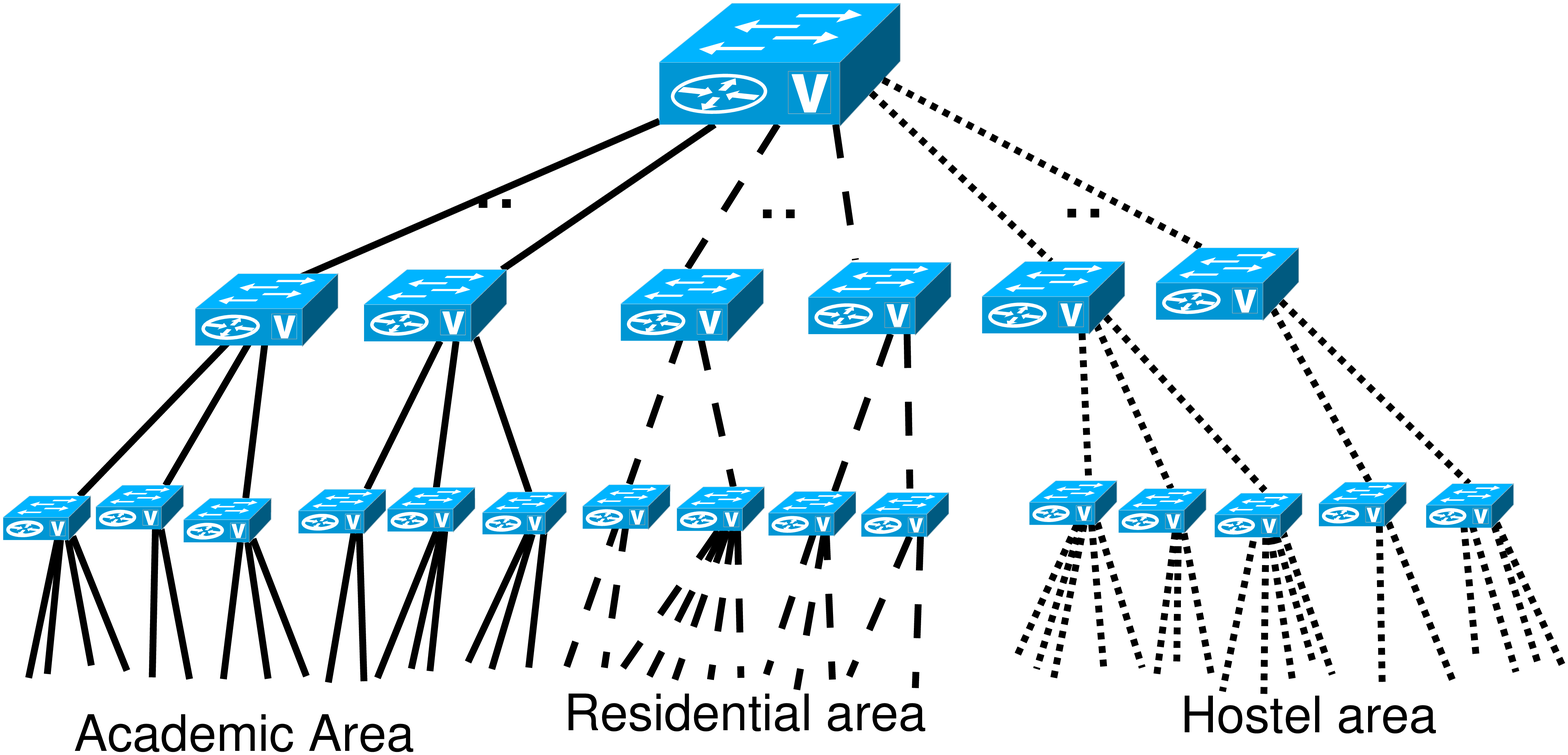}
\caption{Network power consumption as a percentage of total consumption as per the data obtained from competent authorities. Note that IITK Network is ten times greater than LNMIIT Network, both in terms of users and diameter.}\label{fig:division}
\end{minipage}
\hspace{3mm}
\begin{minipage}[b]{0.48\linewidth}\centering
\includegraphics[width=\textwidth]{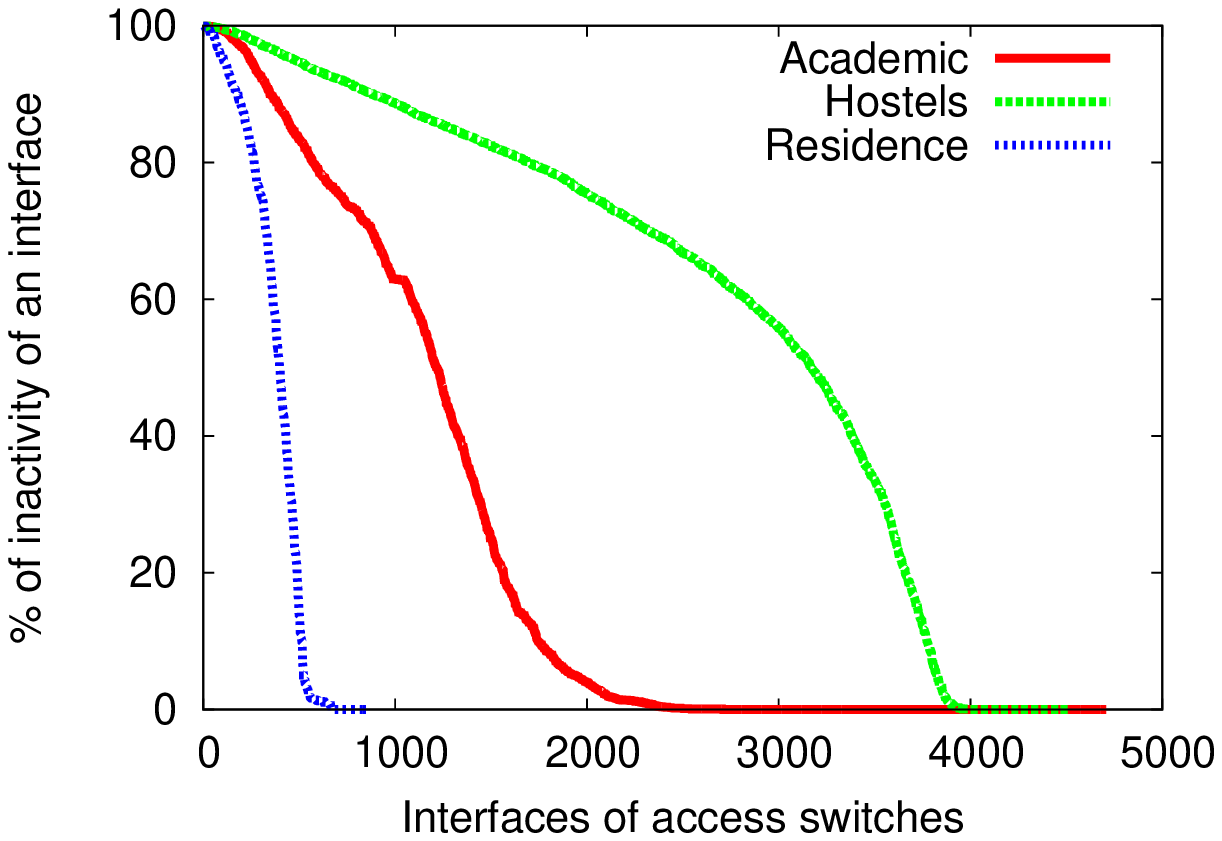}
\caption{Typical hierarchical tree topology of networks in India. The routers are kept only at the edge towards ISP and topology below the core switch is a common characteristic of four networks that shared this information.}\label{fig:data_on_division}
\end{minipage}
\hspace{3mm}
\end{figure*}

We continuously monitored IITK network between $18^{th}$ May to $9^{th}$ July 2014, i.e. for 53 days. It has 1 core switch, 70 distribution switches, and more than 800 access switches. All switches are from just one vendor -- Cisco. Cisco 2960G series is used in the access layer and Cisco 3750X series is used at distribution layer. We connect to each switch and collect five-minute output summary of all the interfaces, every twenty minutes for access switches, and every five minutes for distribution switches issuing the command ``int sh summary" by an automated script. We use data collected from access switch interfaces to find information about the state of the end user terminal. Even if an interface is up without any activity, we consider it as an active user. \deleted {Thus, for us, a user is using the network if the corresponding interface is up.} Since the number of users is huge, for better understanding, we divide IITK Network into into three logical, disjoint networks 1). Academic Area, 2). Faculty and staff residential area, and 3). Student hostels (Fig. \ref{fig:division}). We also collected the switch level topology of LNMIIT Jaipur. 

%

\begin{figure*}[t]
\centering
\hspace{-5mm}
\subfigure[Unused ports at distribution switches are relatively higher in Academic Area, implying that the number of access switches attached to a distribution switch are lower than the number of ports it has.]
{
\includegraphics[width=5.3cm]{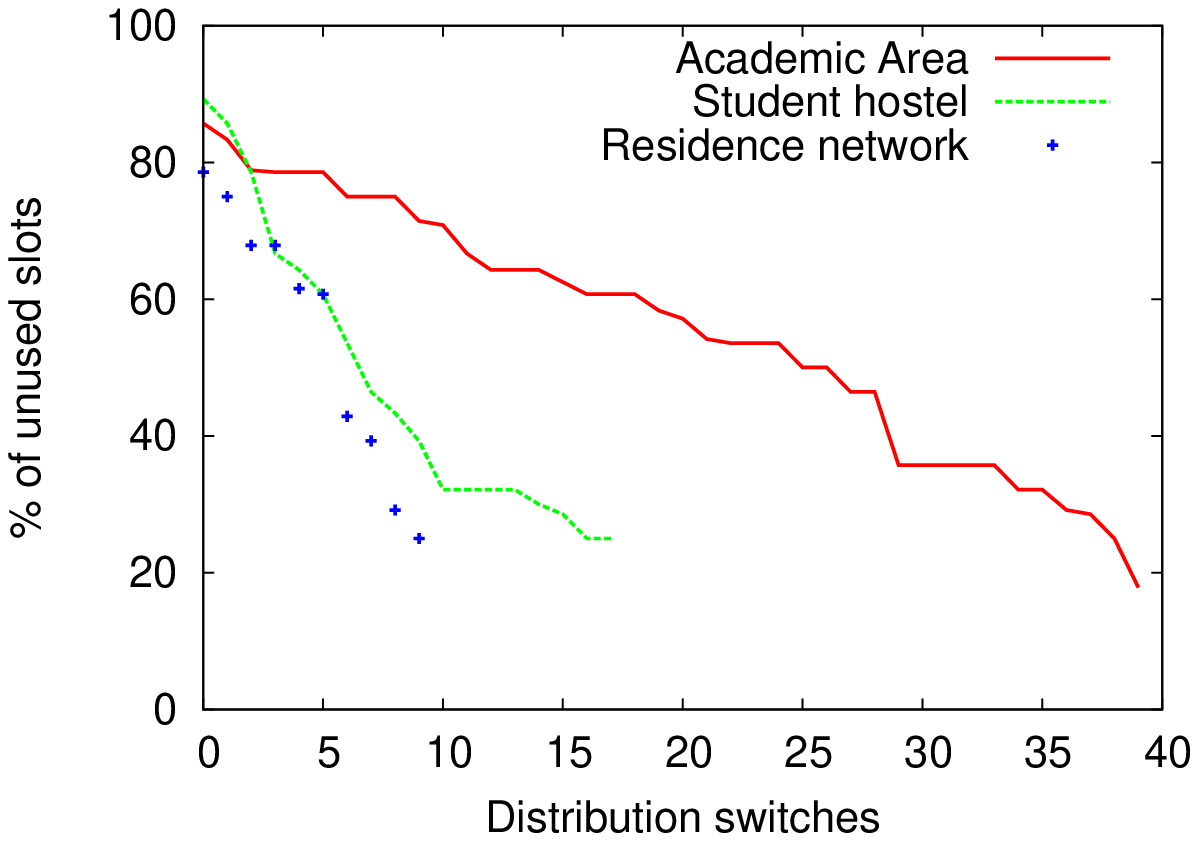}
\label{fig:unused_d}
}
\qquad 
\hspace{-5mm}
\subfigure[Access switches are quite well utilized in residence and hostel area. However, it remains less utilized in Academic Area, implying that lesser number of users are connected to an access switch than its capcity. ]
{
\includegraphics[width=5.3cm]{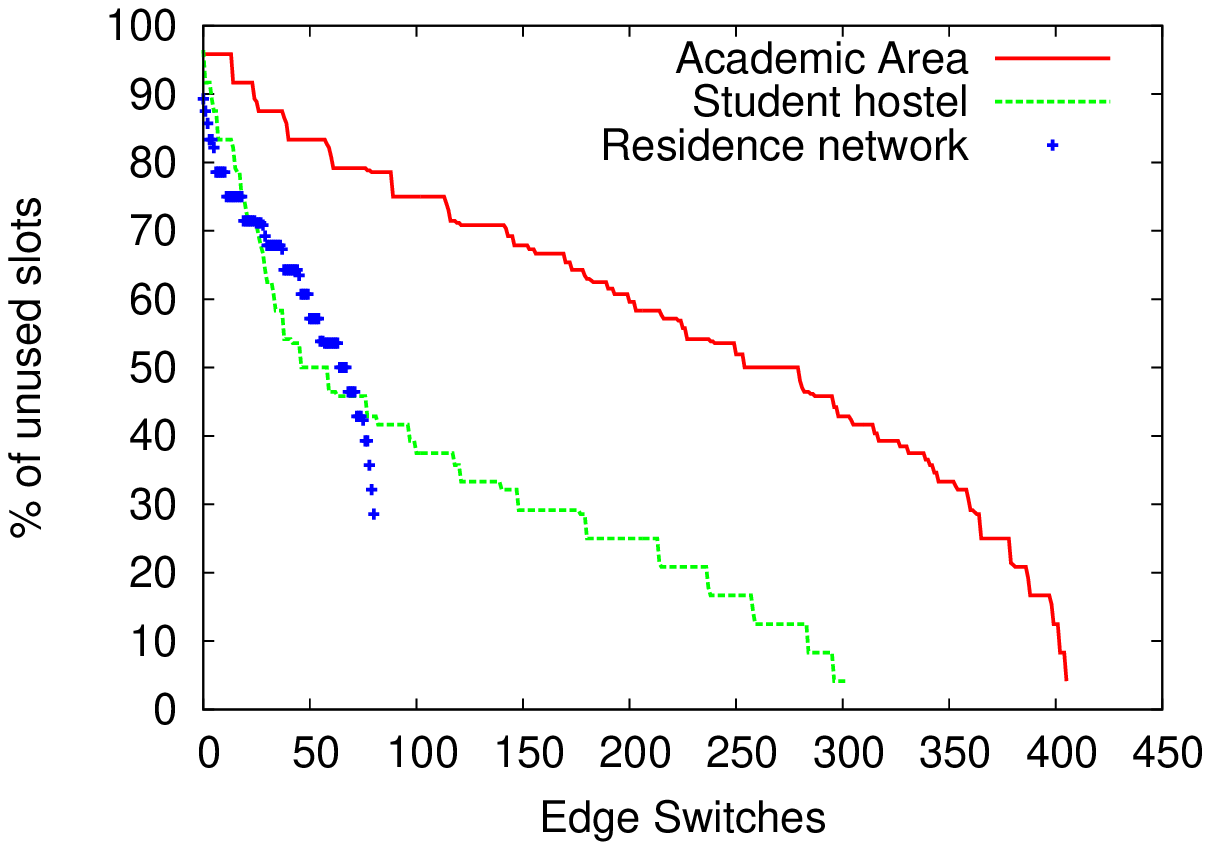}
\label{fig:unused_e}
}
\qquad 
\hspace{-5mm}
\subfigure[Ports of switches in LNMIIT are far more utilized. In fact, it is one of the metric used to minimize the cost of the network. We use values mentioned here to map the number of users to switches.]
{
\includegraphics[width=5.4cm]{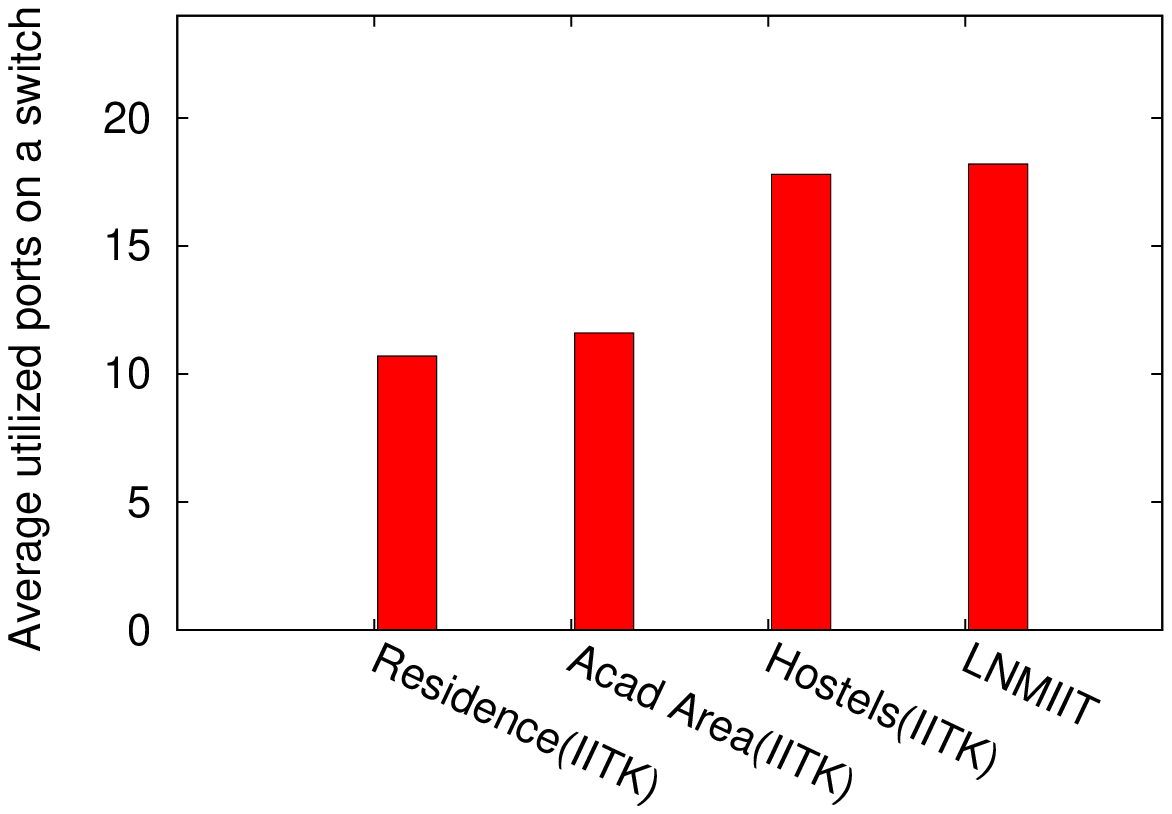}
\label{fig:unused_e}
}
\caption{Potential for cost and power reduction. If all the ports of a switch are utilized on expense of extra cables, fewer switches are enough. In LNMIIT network, high port utilization is already a metric for keeping the cost low. }
\label{fig:cost}
\end{figure*}

\begin{figure*}
\centering
\hspace{2mm}
\subfigure[\% of Knight users in hostels between 23:00-08:00 hours. However, they may come online anytime. Two spikes are due to exams on the next day, resulting in more students leaving the network early.]
{
\includegraphics*[width=5.2cm]{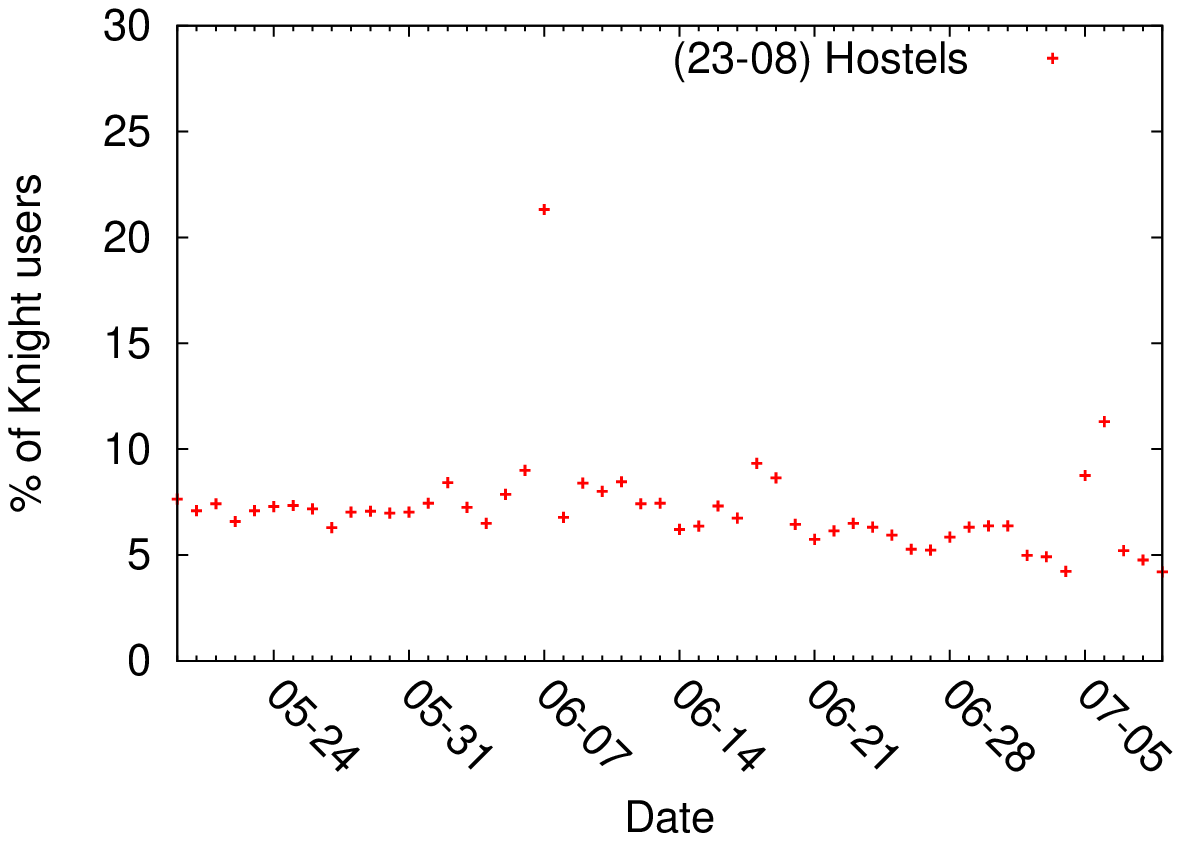}
\label{fig:u0hostels}
}
\hspace{1mm}
\subfigure[Higher \% of Knight users in Residence area between 23:00-07:00 hours as compared to the hostels. Like hostels, users may come online at anytime though.]
{
\includegraphics*[width=5.2cm]{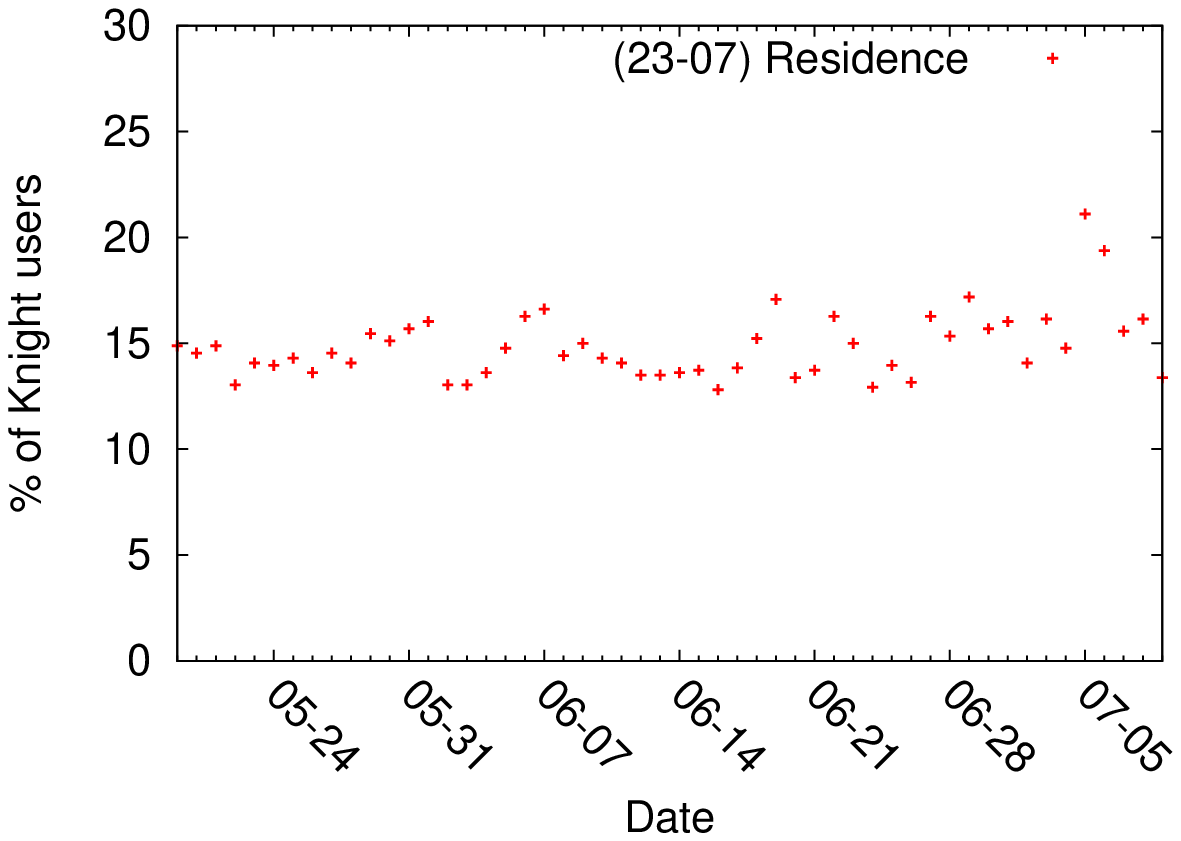}
\label{fig:u0ring}
}
\hspace{1mm}
\subfigure[\% of Knight users in Academic area between 21:00-09:00 hours is quite high. The lined curve represents office users, who are online during day but offline during the night hours.]
{
\includegraphics*[width=5.2cm]{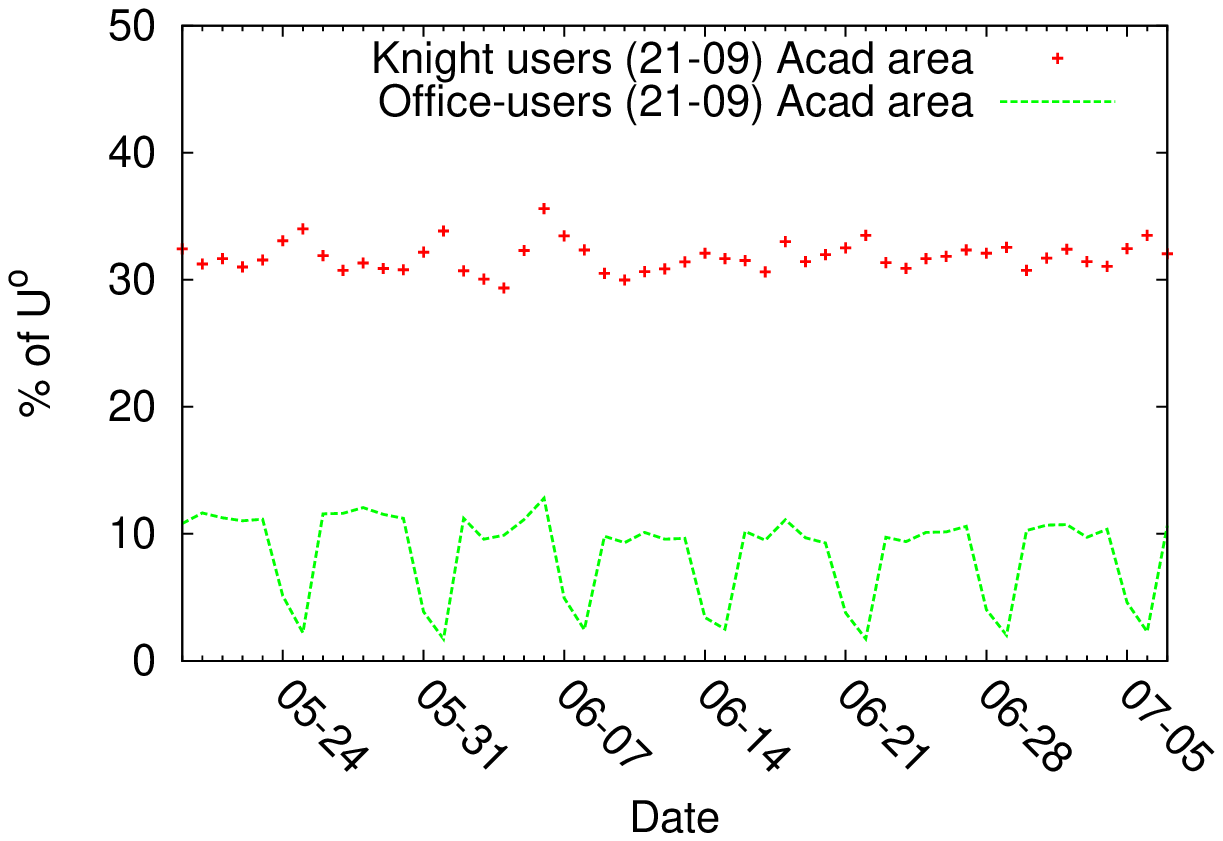}
\label{fig:u0acad}
}
\hspace{1mm}
\subfigure[Number of Knight users on switches in hostels are quite low. Of 18 ports utilized on average (Fig. \ref{fig:cost}(c)), only one of them has 4 Knight users, rest of them have less than four Knight users during 23:00-08:00 hours. ]
{
\includegraphics[width=5.2cm]{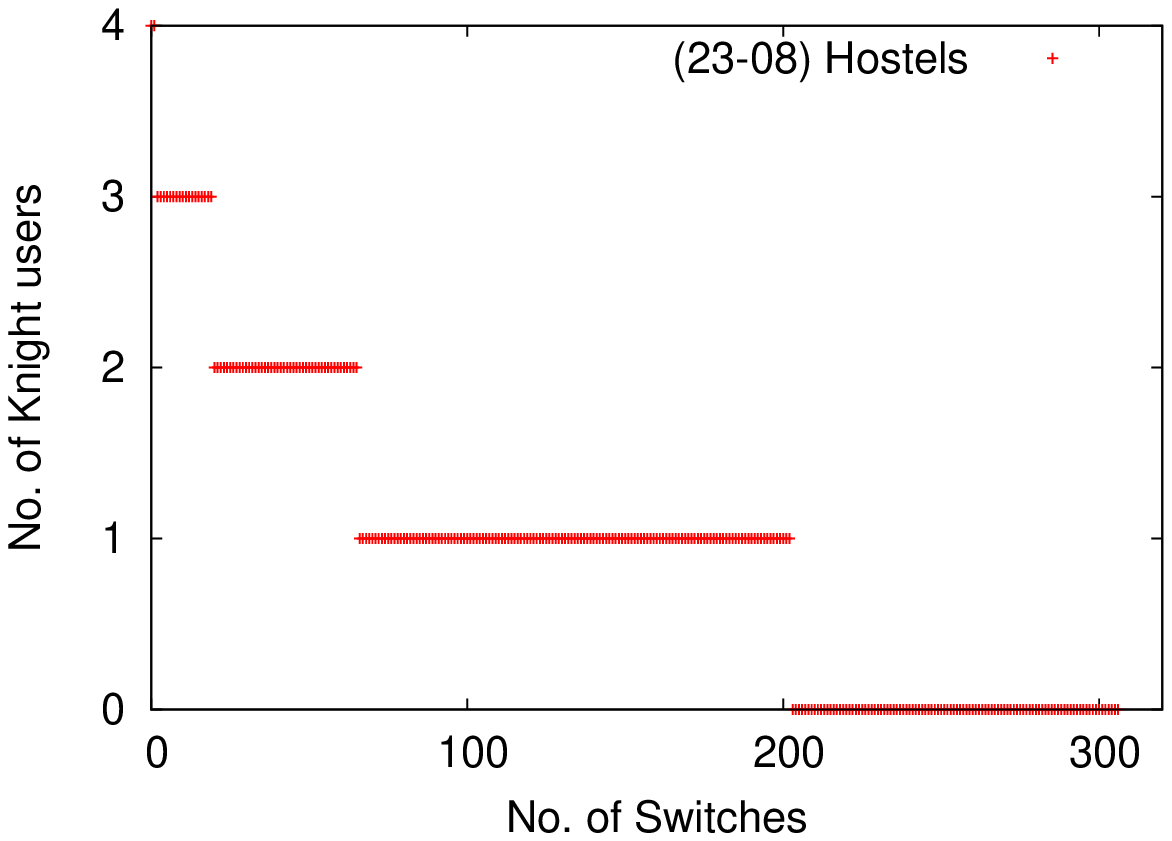}
\label{fig:swthostels}
}
\hspace{1mm}
\subfigure[Number of Knight users between 23:00-07:00 on switches in residential area is higher compared to the hostels. As compared to average utilization of 11 ports (Fig. \ref{fig:cost}(c)), some switches have as high as 10 Knight users.  ]
{
\includegraphics[width=5.2cm]{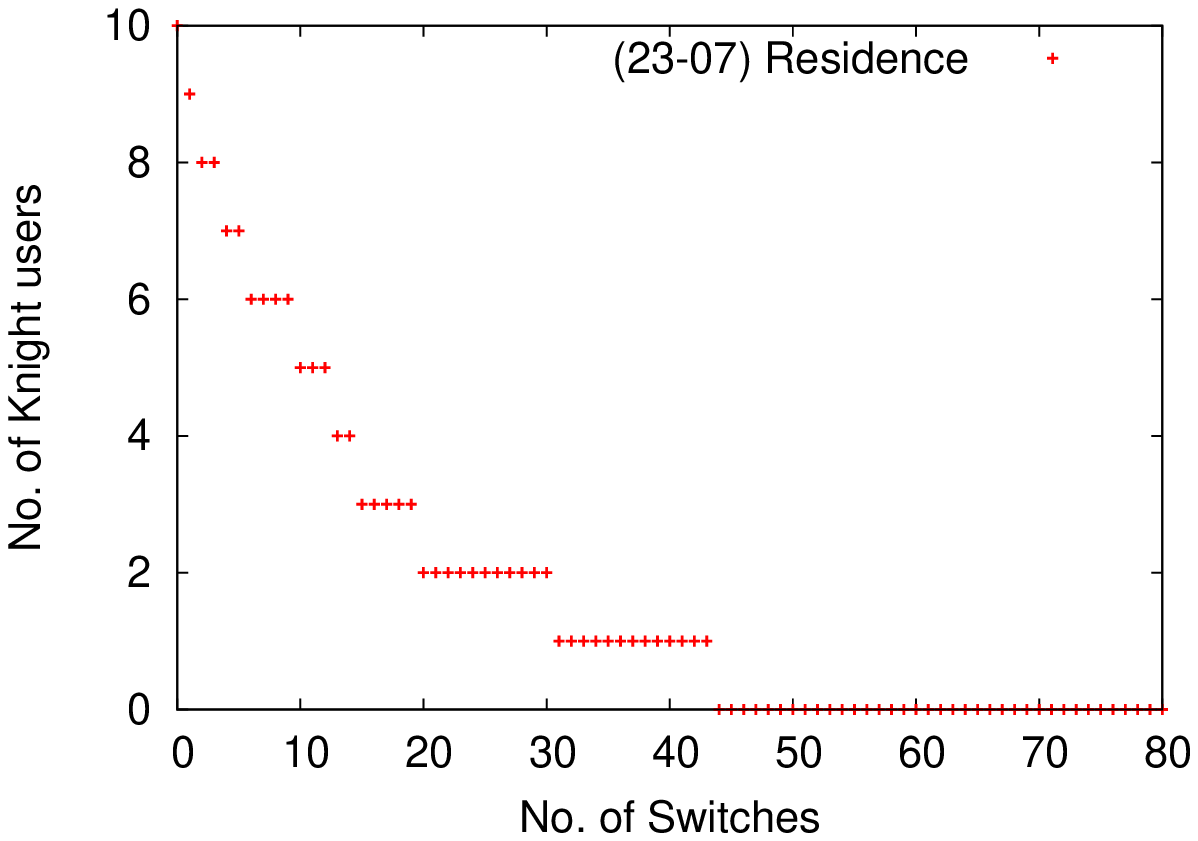}
\label{fig:swtring}
}
\hspace{1mm}
\subfigure[Number of Knight users between 21:00-09:00 hours in academic area switches is comparable to the average (Fig. \ref{fig:cost}(c)). Number of office users is high too, indicating the presence of $\Uo$ users in academic area. ]
{
\includegraphics[width=5.2cm]{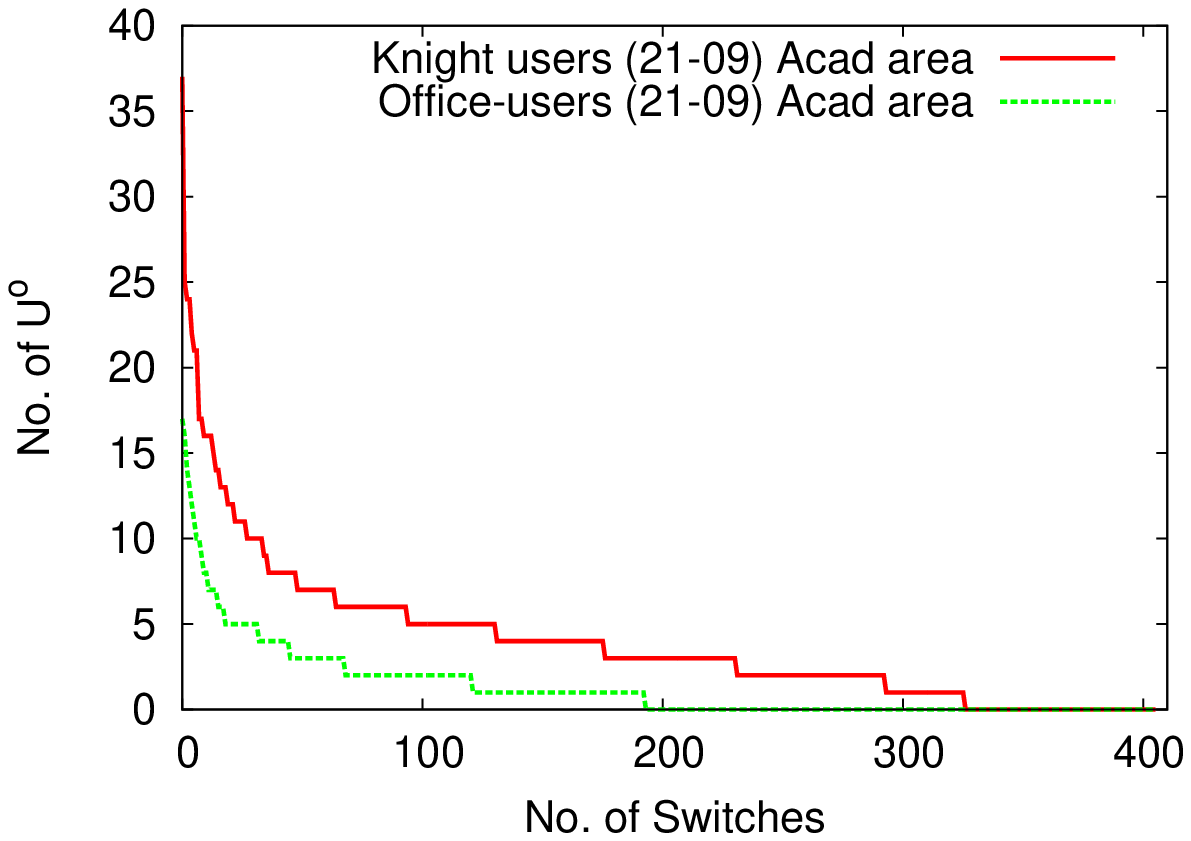}
\label{fig:swtacad}
}
\caption{The figure shows day-wise ((a)-(c)) and switch-wise ((d)-(f)) data of the users. Fig. (a) and (d) are for the hostels. The percentage of users who go offline during night is low and the number of users going offline on any switch is very low. Results for the residential area (Fig. (b) and (e)) improve but not good enough to claim the presence of $\Uo$ users. Fig. (c) and (f) are for academic area and they clearly indicate the potential candidates for being $\Uo$ users and $\Ao$ switches. There are substantial number of users who go offline during night and on weekends. Also there are substantial number of switches that have high number of $\Uo$ users.
}
\label{fig:officeusers}
\end{figure*}

\begin{figure*}
\centering
\subfigure[Number of Knight and office users in different buildings of Academic Area. If some access switches are exclusively connected to these users, then they can be switched off during night. ]
{
\includegraphics[width=5.5cm]{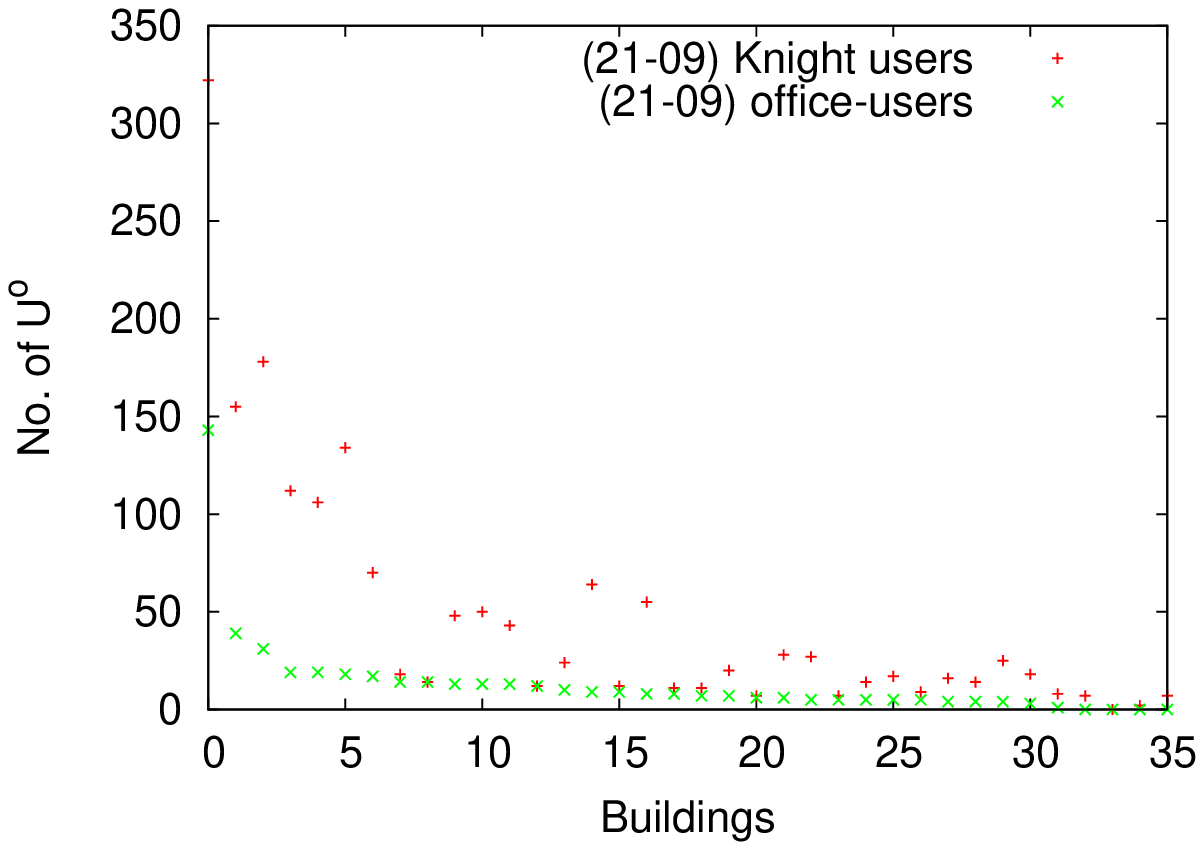}
\label{fig:buildingwise}
}
\hspace{1mm}
\subfigure[We have a switch level topology of the network. Assuming each switch is connected to 12 users (Fig. \ref{fig:cost}(c)), number of $\Ao$ switches in each building is obtained, which in turn, gives an estimate on energy savings.]
{
\includegraphics[width=5.5cm]{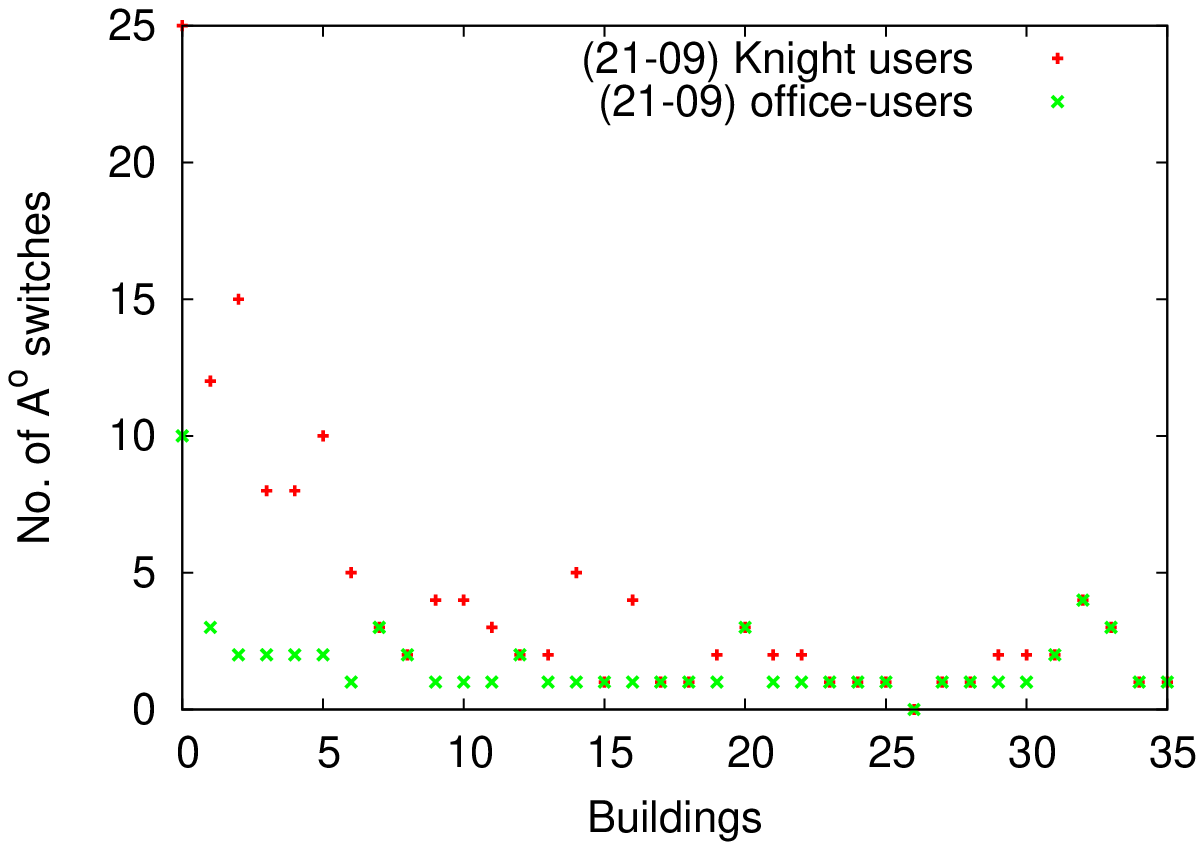}
\label{fig:building_switch}
}
\hspace{1mm}
\subfigure[Energy savings for whole IITK network and just academic area is shown. It is clearly visible that in office only environments, substantial energy savings are possible.]
{
\includegraphics[angle=270, origin=c, width=4.7cm]{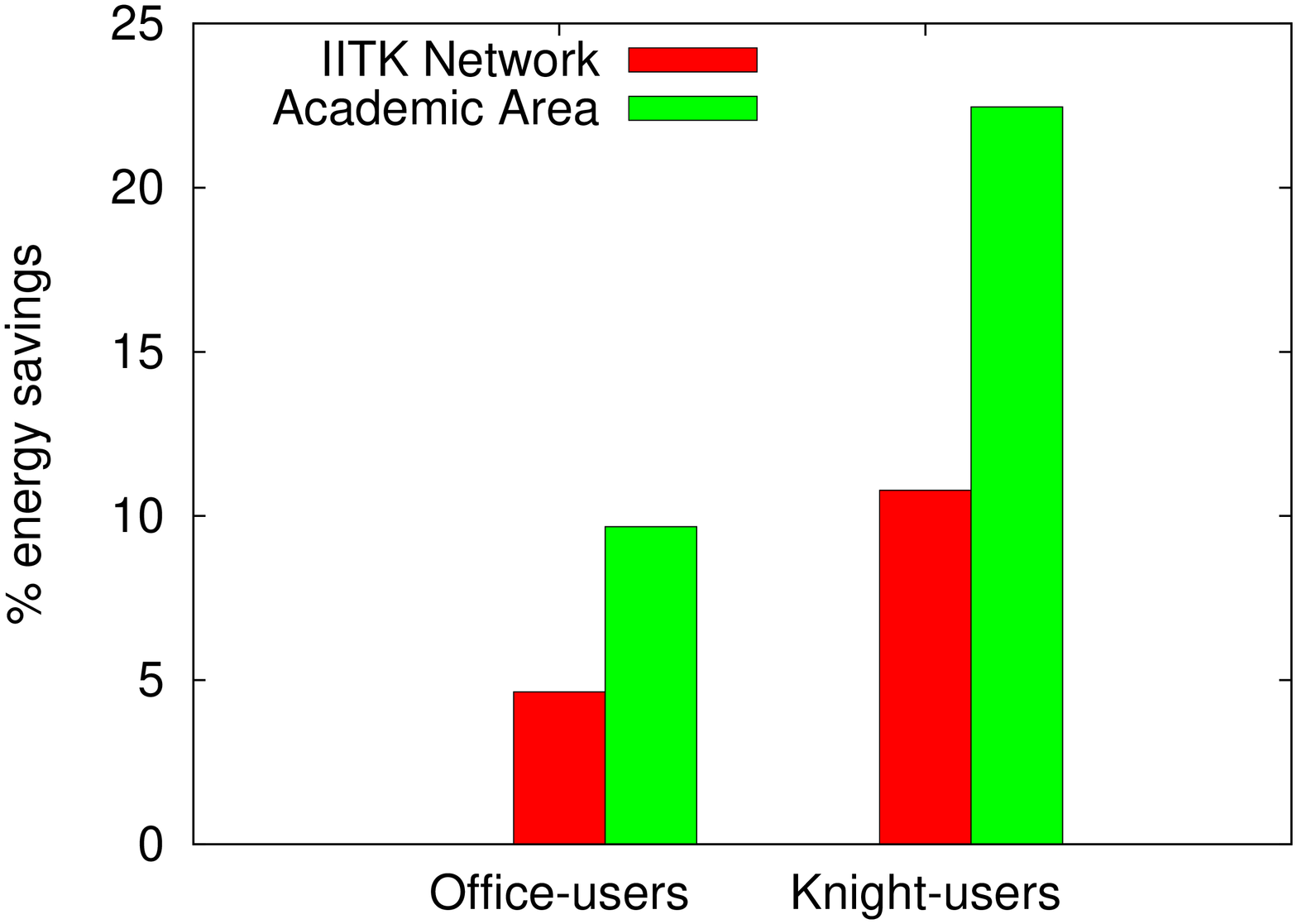}
\label{fig:energy_savings}
}
\caption{Subfig. (a) shows the number of $\Uo$ in each building of academic area which is converted to the number of $\Ao$ using Fig. \ref{fig:cost}(c). Finally (c) shows 4.64\% and 10.78\% reduction in the monthly power consumption of IITK-Network by office-users and Knight users respectively. Energy savings of just academic area up to 9.67\% and 22.46\% for office-users and Knight users respectively.}
\label{fig:officeswitches}
\end{figure*}

\subsubsection{Data on end users}
\deleted {The network serves to 350 faculty members, 900 staff and 4000 students.} Fig. \ref{fig:data_on_division} gives an estimate of the number of interfaces that were up at least once during the whole data collection period. In the academic area more than 4500 interfaces were used at least once during the data collection period. Similarly, in the student hostels more than 4500 interfaces and in the residential area more than 800 interfaces of the access switches were active at least once. We emphasize that if an interface has been up for even once in the collected data, we have included it in the figure. It is for this reason there are many interfaces with inactive percentage close to 100 (but none exactly 100). From the figure, it can be seen that more than two-third of the hostel interfaces and half of the residential interfaces remain inactive for more than 50\% of the time, whereas in the academic area only one-third of the interfaces are inactive for more than 50\% of the times. The academic area also has the maximum number of users with 100\% activity. This is due to the connections in data center, switches to which the wireless access points are connected and other computing devices that are up and running round the clock throughout the year. \deleted{From now on, we use the terms user and interface interchangeably unless noted explicitly.}
\subsubsection{Utilization of switches}
Fig. \ref{fig:cost} shows the utilization of ports of switches at different levels in IITK network. It also shows the average number of ports utilized in different networks. High utilization tends to justify the fact that the universities are tight on budget. So they use as fewer switches as they can compared to low port utilization at an enterprise network in a developed country \cite{mahadevan2010energy}.

\subsection{Inferring users in $\Ut$ and $\Uo$}
\label{subsec:inferring_users}

\added{After careful examination of data, we are able to find substantial number of $\Uo$ users that enable energy savings.} Recall that $\Ut$ users are the ones who may require access to the network anytime. It is not desirable to turn off switches connected to such users. On the other hand, $\Uo$ users have a schedule in which they use the network. From Fig. \ref{fig:data_on_division} it seems that the hostels should admit the maximum number of $\Uo$ users because of the maximum number of interfaces with high inactivity. However, inactivity percentage does not directly predict the type of users because it does not imply the schedule of the users. Users in the hostels may come and leave the network at will. Hence, we cannot turn off any of the switches in the hostels. \deleted{as network availability for twenty-four hours of a day is important.} Similarly, for residential users, the network should be available twenty-four hours. In academic area, even though the percentage of inactive interfaces is relatively lower compared to hostels and residence, it should admit $\Uo$ as the users who leave the network after work. In India, it is strongly recommended to switch off electrical equipments when the employees leave offices, including PCs. \deleted{assume that 1500 (post-graduate) students  and 350 faculty members may require access to the network any time. There are switches connected to devices like wireless access points, servers, etc. Around 400-600 users who have a fixed working hours can be put in $\Uo$ category.} The question that now arises is what if the users belonging to $\Uo$ category want to access network from their home? It turns out that there are users that never access their computers once they leave the office.

We look for a similar conclusion in the data that we have collected. \added{Before we proceed further, we introduce two kinds of users that can be offline during the night, defined below:}

\begin{defi}
\emph{(Knight users): Users that are offline during the night hours but may or may not be using the network during the day. }
\end{defi}

\added{
\begin{defi}
\emph{(Office users): Users that are offline during the night hours but are online at least once during the day on a working day. }
\end{defi}
}

Note that the Office users are a subset of Knight users. Figures \ref{fig:officeusers}(a)-(c) show the percentage of \emph{Knight} users in the three logical divisions of the network.

\subsubsection{Hostels and residence belong to $\Ut$}
\label{subsubsec:hostel-resi-ut}
In Fig. \ref{fig:officeusers}(a), we fix night hours between 23:00:00-07:59:59, i.e., users should never be using the network between these hours but may or may not use it during the day. This results in roughly 7\% \emph{Knight users}. The problem is that they are spread over more than 4000 users located in different hostels and are not necessarily the same interfaces every day. \replaced{Moreover, the hostel users may require access to the Internet anytime, the availability of the network in such areas is crucial.  So they cannot be put into $\Uo$ category. }{ Even if they are the same, availability of the network to a user who may use it is crucial. } Hence, users in hostels belong to $\Ut$. The hike on the night befored $7^{th}$ June, and $9^{th}$ July is due to the examinations of summer term on those days. So more students went off the network before 23:00. \deleted{ and hence this abrupt increase.} Fig. \ref{fig:officeusers}(b) shows similar data for the residential network. Even though the percentage of users that don't use the network in the night increases, \deleted{the interfaces in the curve may not be the same all the time. Even if they are the same,} network in the residential area cannot be turned off for the reasons similar to hostels.

\subsubsection{Academic Area accommodates some $\Uo$ users}
\label{subsubsec:acad-uo}
Fig. \ref{fig:officeusers}(c) shows interesting data with more than 30\% of the \emph{Knight} users for night hours taken to be 21:00-09:00. Only those users show up in this curve that appear at least for 3 days (to make sure that it is not just a one time affair) in the collected data. We do a further analysis of this segment of the network. The other curve in this figure represent the percentage of users that are offline during the night hours, online at least once during the day, and appear in at least 50\% of the working days in the data collection period. Such users, with high confidence, can be classified into $\Uo$ category. It can be seen that there is a sharp periodic decrease in the percentage for two days. A careful observation tells us that these two days are weekends and the reason for drop in the percentage is the fact that we capture ONLY those users that are ON at least once in the day and OFF during the night hours. During weekends, it does not hold, as most of the users are not ON during the day. Hence, such users do not appear. \deleted{This gives us a strong motivation to evaluate energy savings that are possible through this design.}

\subsection{Finding $\Ao$ switches in buildings in Academic Area}
\label{subsec:finding_switches}
Our goal is to be able to find the switches that can be turned off during the mentioned night hours. Figures \ref{fig:officeusers}(d)-(f) show the access switch data for the network. As discussed before, the day users in hostels are very much spread out, which is confirmed by Fig. \ref{fig:officeusers}(d) for hostels. No switch has enough \emph{Knight} users. Even though in the residential network, some switches have relatively good number of \emph{Knight} users (Fig. \ref{fig:officeusers}(e)) but no switch of this network can be turned off.

Now we discuss \ref{fig:officeusers}(c) and \ref{fig:officeusers}(f) in relation with each other. We consider both ``office-users" and \emph{Knight} users to be potential candidates for $\Uo$ and do further evaluation keeping this in mind. We observe that there are switches with high number of $\Uo$ users. 
\deleted{Now that we have potential candidates to be in $\Ao$, we still need to apply the length constraint described in Eq. \ref{eq:c7} which we have restricted to the same building in Section \ref{sec:approach}.} From Fig. \ref{fig:officeusers}(f), we just add up the number of $\Uo$ users of switches that belong to the same building to obtain Fig. \ref{fig:officeswitches}(a). Recall that we have switch level topology of IITK-Network. Fig. \ref{fig:officeswitches}(a) is used to obtain Fig. \ref{fig:officeswitches}(b) by counting a switch equivalent to 12 users. 12 users per switch is an average for the academic area of IITK-Network (Fig. \ref{fig:cost}(c)). It is expected that set of office-users is contained in the set of \emph{Knight} users but we see that they are exactly the same for three buildings. On other hand, in Fig. \ref{fig:officeswitches}(b) we see that $\Ao$ is same for 18 buildings implying that for most of the buildings office-users already capture most of the $\Uo$ users. We find that there are four buildings for which the number of office-users is zero but only one of them is zero for \emph{Knight} users. A little bit more analysis had revealed that those buildings close between 21:00 hours to 00:00 hours. Hence, the interfaces from these buildings never occurred in the office-users data. However, some interfaces may still appear in \emph{Knight} users. For example, an interface that was unused on a particular day.

\subsection{Energy savings}
\label{subsec:energy_savings}
Based on the analysis done until now, Fig. \ref{fig:officeswitches}(c) briefs the energy savings using the technique presented in the paper by switching off the switches of type $\Ao$ during night-hours and weekends. Hence, the result is obtained as percent monthly savings from Fig. \ref{fig:officeswitches}(b) using the following formula:
\[ \% = \frac{\#\Ao \times (night hours \times working days + 24 \times 8 )}{\#\A \times 24 \times 30}   \]

Number of $\Ao$ turns out to be 62 for office-users and 144 for \emph{Knight} users for academic area that has 406 switches and the whole network has 846.   This implies 4.64\% and 10.78\% reduction in the monthly power consumption of IITK-Network respectively. Whereas the energy savings of just academic area  network shoots up to 9.67\% and 22.46\% respectively. 

In our approach, the length of wires between the user and edge switch can increase. Assuming this to be 30m per user (maximum width of a normal building). There are 476 $\Uo$ users, so 14,280m (maximum) of extra wires to be needed to implement our idea for saving 4.64\% (the minimum energy savings) energy per month. Total cost invested in wires would be Rs. 214,200 (note that all the wires used in the buildings are made up of copper, see Table \ref{tab:n1}). The electricity bill saving through our method is Rs. 18,560 per month. Thus the cost invested can be recovered by saving in electricity bill within 11.54 $<$ 12 months.

\begin{figure*}[t]
\centering
\subfigure[$\Ao$ kinds of switches that are physically outside but in the network topology they are in academic area network. ]
{
\includegraphics[trim=40mm 105mm 20mm 73mm, clip, width=5.7cm]{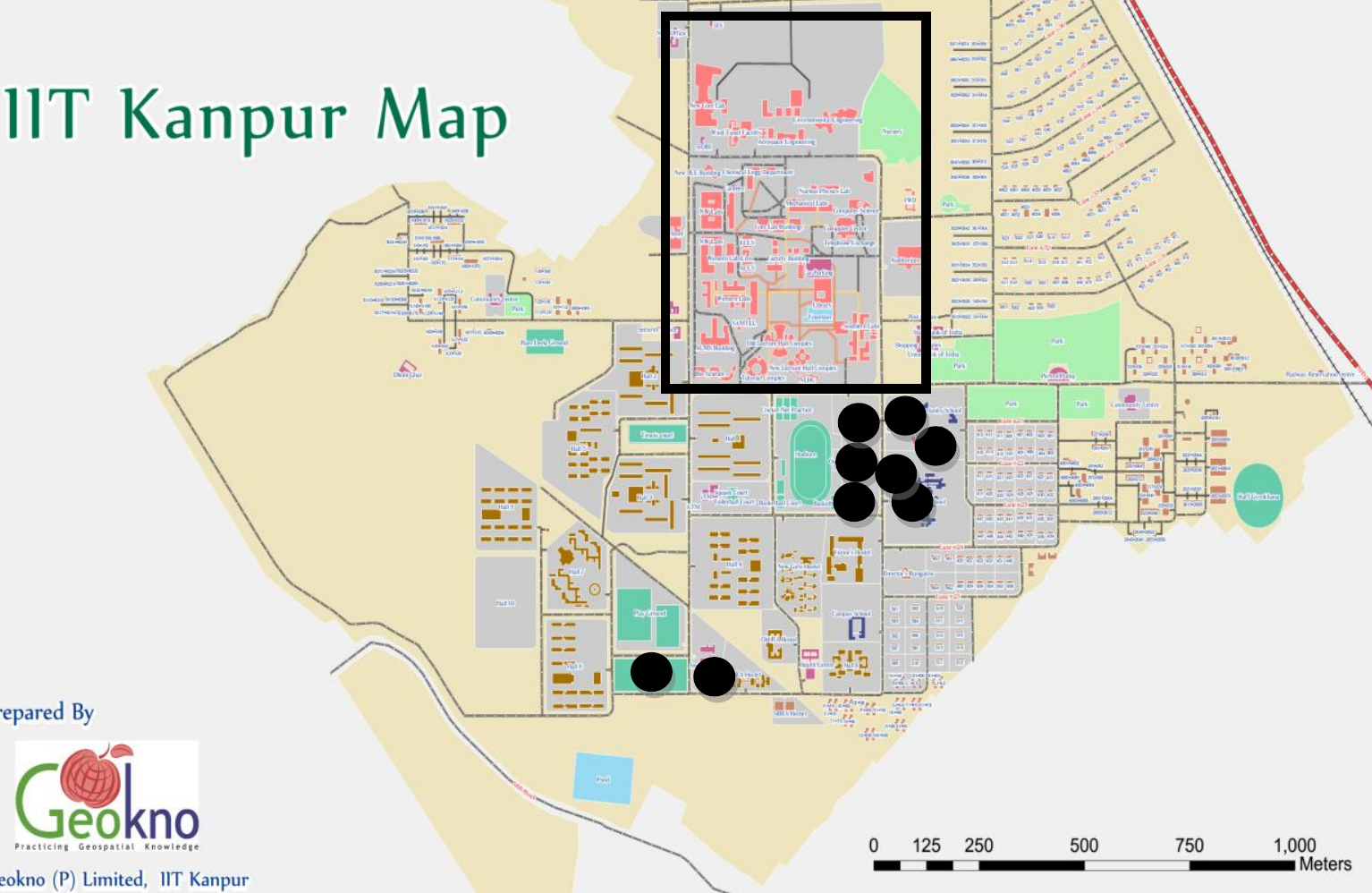}
\label{fig:n1_dotted}
}
\qquad
\subfigure[$\Ao$ kinds of switches physically as well as logically inside the academic area. ]
{
\includegraphics[trim=10mm 120mm 15mm 35mm, clip, width=5.7cm]{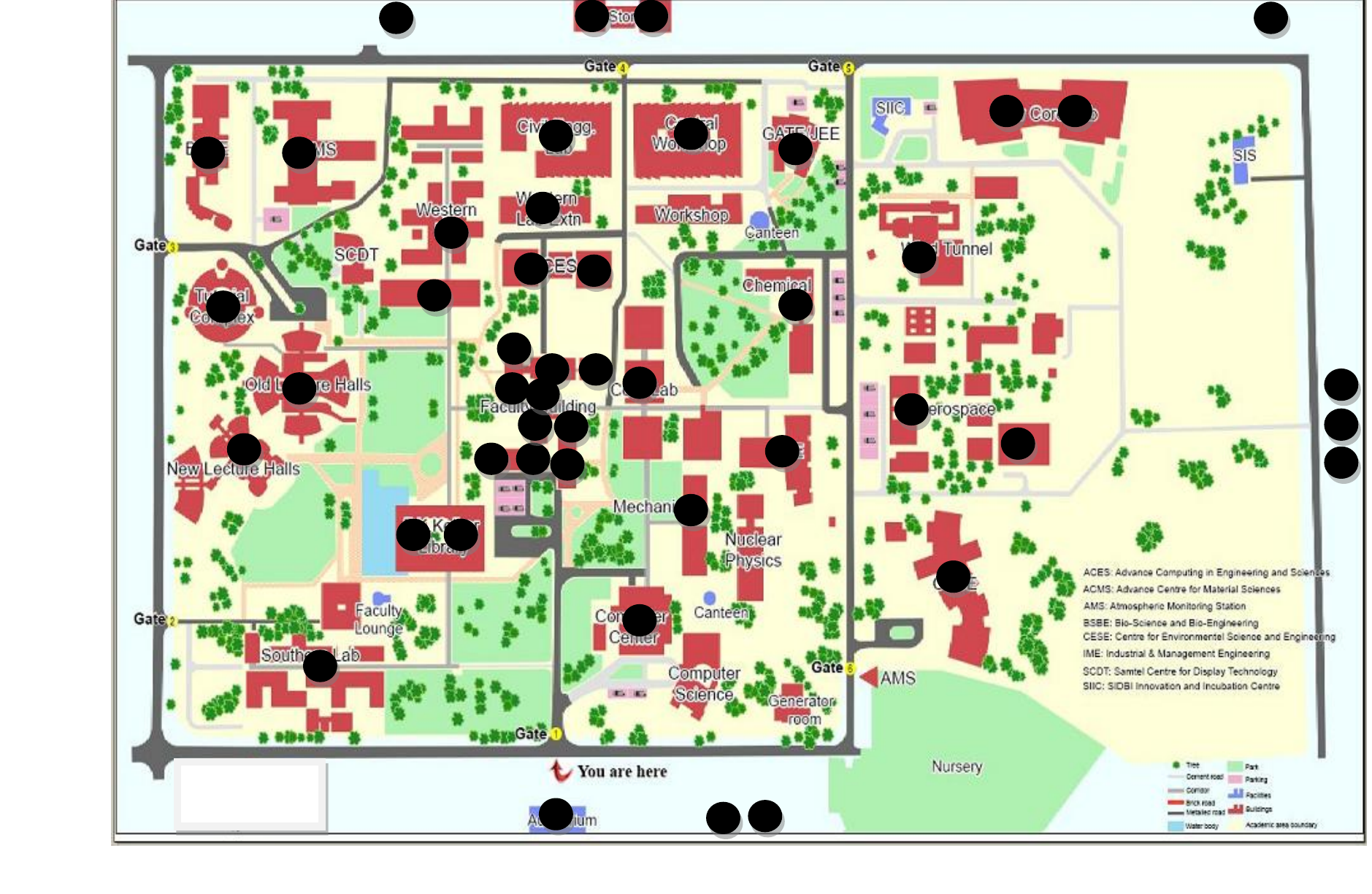}
\label{fig:acad_dotted}
}
\qquad
\subfigure[Each user puts a ``x" against their name if their PC is switched off. The last user looks for all ``x".]
{
\includegraphics[width=2.8cm]{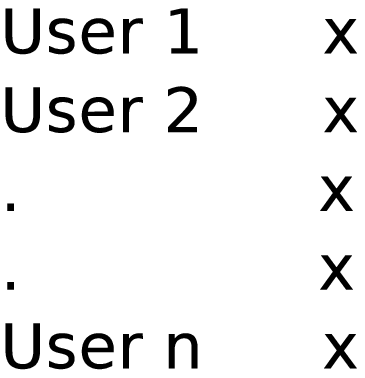}
\label{fig:acad_dotted}
}
\caption{The above two figures show the location of switches that can be turned off during office hours. The rectangle marked in Fig. \ref{fig:maps}(a) has been rotated clockwise 90 degrees and expanded as shown in Fig. \ref{fig:maps}(b), represents the physical boundaries of the academic area. In total, 62 switches can be switched off during 21:00-09:00 hours, off a total of 846. For better view, please refer to the detailed map of IIT Kanpur \cite{iitkmap} \cite{acadareamap}.}
\label{fig:maps}
\end{figure*}

\section{Implementation}
\label{sec:implementation}

Having talked about the potential energy savings through this method, we are still left with one question unanswered, i.e., who turns off the switches? It is a common practice in India to put labels like ``If you are the last one to leave, switch off all the electrical equipments" on the walls of the offices. Hence, we propose that the last user of the network turns off the switch before leaving. Note that since office users are recommended to shut down their PCs before leaving, such switching off of the switches should not cause any loss to any other user. 

However, assuming no loss to any user may be dangerous. So the first step is to spread an awareness among the users about energy savings through network. To begin with, a paper can be put on the notice board besides the power button for switches (Fig. \ref{fig:maps}(c)). Everyone connected to a particular $\Ao$ switch checks it before leaving the office if their PC is turned off. If the last user of this switch sees ticks from all users, he/she turns switch power off before leaving the office premises. In the morning, the first user can turn it on again. Such a distributed implementation has an advantage over the centralized implementation of switching off the switches by CC staff. It saves time as it may take up to an hour for one single person to switch off all the switches. Moreover, such an aggressive power saving may trigger irresponsible users to save power from other sources as well.

\begin{table*}[t]
\begin{center}
\begin{tabular*}{173mm}{|c|c|c|c|c|c|p{18mm}|}
\hline
\textbf{Vendor} & \textbf{Model} & \textbf{Power$_{24}$(W)} & 
\textbf{Price$_{24}$($\$$)} & \textbf{Power$_{48}$(W)} & \textbf{Price$_{48}$($\$$)} & 
\textbf{Op.Temp.} \\
 \hline\hline
Juniper  & EX2200-24T-4G & 50 & 1,212.80 & 76 & 2044.80 & $0-45\,^{\circ}\mathrm{C}$   \\ \hline
Cisco  & WS-C2960G-24TC-L & 55 & 2,277.0 & 80 & 4137.00 & $0-45\,^{\circ}\mathrm{C}$  \\ \hline
D-Link  & DGS-3420-28TC & 50.8 & 1,529.99 & 81 & 2549.99 &$0-50\,^{\circ}\mathrm{C}$   \\ \hline
HP & HP 2920-24G (J9726A) & 58 & 1065.60 & 70 & 1873.85 & $0-55\,^{\circ}\mathrm{C}$    \\ \hline
\end{tabular*}
\end{center}
\caption{Some of the available switches for access layer.}
\label{tab:iitk}
\normalsize
\end{table*}

\section{Pinging in the network}
\label{sec:pinging}

\begin{figure}[t]
\centering
\includegraphics[width=12cm]{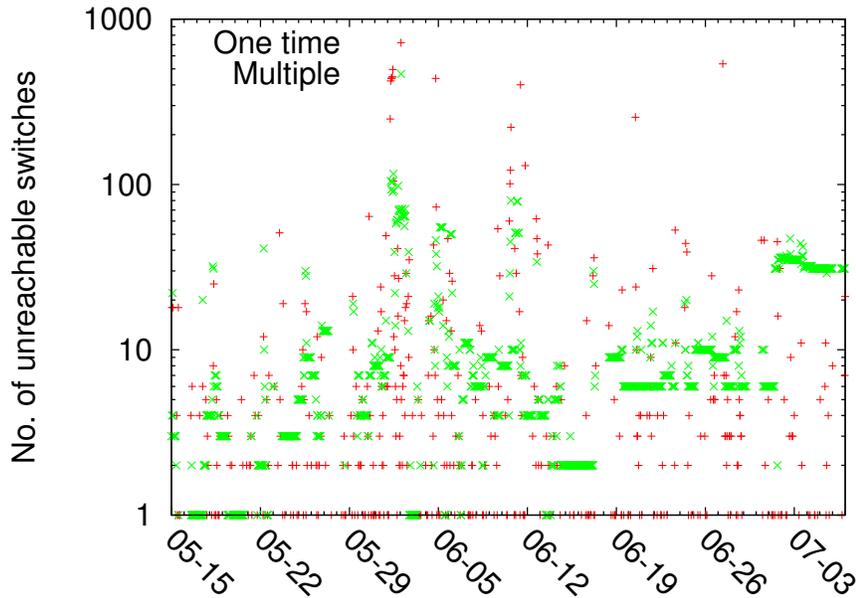}
\caption{Number of unreachable switches during different dates, plotted hourly. A `+' represents the number of switches which did not echo ping request exactly once whereas a `$\times$' represents the number of switches that did not echo ping request more than once in the considered hour.}
\label{fig:unreachable}
\end{figure}

We ping every switch in the network for 55 days starting from $15^{th}$ May to $8^{th}$ July 2014, every 20 minutes, the minimum time period allowed by the computer center staff for continuous pings. Fig. \ref{fig:unreachable} presents the hourly result of ping experiment collected. We divide the failure of ping in two cases, 1). When the ping request is not echoed exactly once in an hour, 2). When it is not echoed more than once. The idea is to capture long term ans short term failures which can happen due to various reasons like a). Maintenance b). Power outages c). congestion d). ambient temperature. If the computer center (CC) staff turns off a distribution switch for maintenance, that often leads to 20-25 switches being unreachable. As mentioned by CC staff, this happens regularly on a daily basis for different buildings. Sometimes because of power outage in a building, all the switches in the building get switched off (this may account for up to 20-25 switches) and is not a rare event. Thus long durations of failures are mainly due to the first two kinds of reasons. 

One time failures are effected due to rise in ambient temperature, switches reboot frequently they are kept without any cooling whatsoever. We did an analysis of such failures and it turns out that many of such failures were during afternoon 12:00-15:00 hours. It is bound to happen when the ambient temperature is unusually high, the highest ambient temperature this summer was $47\,^{\circ}\mathrm{C}$, which is beyond the operational temperature of these switches \cite{2960datasheet}. According to the computer center staff, network becomes crazy during summer. They have to keep running all the time to fix the faults whereas their job is much easier during winters. Congestion level of the network is high but we don't comment much on it in the interest of space. 

In Table \ref{tab:iitk}, we compare some of the available switches that can be used at the access layer of a network. We see that power consumption is more or less the same for all the switches whereas the cost varies drastically. All the prices are quoted from the latest editions of PEPPM \cite{peppm}. We also observe that switches from some vendors have higher operating temperature range, which is beneficial in our case. Hence, choosing all D-Link or HP switches may lead to cheaper networks with better quality of service in summer. A similar comparison of switches at distribution layer is possible. Note that we are considering only 24 port switches, a similar comparison for 48 port switches can be done.

\section{Related work} 
\label{sec:relatedwork}

The need for energy efficiency and its importance in developing countries was first studied in seminal paper by Gupta and Singh \cite{gupta2003greening}. Since then research for designing energy efficient networks have been one of the major research drives and many important references can be found in \cite{bolla2011}. A key observation in the solutions is to exploit the over-provisioning in the networks. Most of the research proposing energy efficiency implicitly considers only the networks in the developed countries as they assume presence of redundant links and/or nodes in the network \cite{mahadevan2010energy} \cite{mahadevan2009energy} \cite{chabarek2008power}. A good amount of literature exists in greening the protocols used for dynamic routing like OSPF \cite{cianfrani2012ospf} \cite{amaldi2011energy}\cite{bianzino2012grida}.

Our work is different from most of these solutions because we provide energy saving network design for a network that 1). has tree topology, i.e., without any redundancy and 2) deploys static routing and 3) is an important network in a developing country like India. Deployment of networks in developing countries pose many challenges \cite{nungu2011powering}. Like us, \cite{he2012greenvlan} study designing of green VLANs but their approach is to make use of the energy efficient hardware, whereas our approach is general enough to be applied to any network to obtain energy savings. \cite{jardosh2009green} propose network on demand approach but for wireless LANs. Switching off of switches by migrating inactive users to wireless network has been proposed in \cite{le2010performance}.

Energy savings in the access network is a difficult task and many innovative methods like \cite{NikosDSLAMSigcom}\cite{reich2010sleepless}\cite{allman2007enabling} have been proposed. A relevant reference for access network design that is close to our design is in \cite{andrews1998access}\cite{gollowitzer2011two}\cite{rodriguez2009improved}  and the references therein.

\section{Conclusion and Future Work} 
\label{sec:conclusion}
Our main observation is the existence of users that allow switching off the networking devices during night hours. However, it is not straightforward to implement the energy savings as the network topology is a tree in which any switched off node may imply disconnection of the users. The issue of electricity saving in the network has to be taken as seriously as for other appliances. We also provide a way of making sure that a switch is not turned off while it is being used by some user. We have also observed that due to extraordinary hot summers, switches that have higher operating temperature range should be used to minimize the reboot of switches due to heat. Our next goal is to do a similar study for a network that uses routers and dynamic routing of packets. We expect more energy savings in such an environment as there are more redundant paths.

\section*{Acknowledgements}
The first author would like to thank Mr. Navpreet Singh, Navneet Sharma, Vijay Gaur and the whole network group at Computer Center, IIT Kanpur for their help with data collection and providing all the necessary and important information related to the LAN of IIT Kanpur.

\bibliographystyle{plain}
\bibliography{paper}

\end{document}